\documentclass[11pt,a4paper]{article}

\RequirePackage{ifpdf} 
\usepackage{amsmath} 
\usepackage{mathtools}

\usepackage{jheppub}
\usepackage{pstricks}
\usepackage[final]{pdfpages} 
\usepackage{ifpdf} 
\usepackage{slashed}
\usepackage{bbm}

\usepackage{hyperref}
\usepackage{cleveref}

\usepackage{color} 
\usepackage{graphics}

\usepackage{etoolbox} 
\usepackage{fixmath}

\usepackage{notoccite} 

\usepackage{caption} 
\usepackage{subcaption} 
\usepackage{amsfonts}

\usepackage{multirow}
\usepackage{epstopdf}

\usepackage{relsize}
\usepackage{cancel}

\usepackage[normalem]{ulem}

\usepackage{tikz}
\usetikzlibrary{positioning,arrows}
\usetikzlibrary{decorations.pathmorphing}
\usetikzlibrary{decorations.markings}
\usetikzlibrary{shapes.geometric}
\tikzset{
    vector/.style={decorate, decoration={snake}, draw},
    provector/.style={decorate, decoration={snake,amplitude=2.5pt}, draw},
    antivector/.style={decorate, decoration={snake,amplitude=-2.5pt}, draw},
    fermion/.style={draw=black,
      postaction={decorate},decoration={markings,mark=at position .55
        with {\arrow[draw=black]{>}}}}, 
    fermionbar/.style={draw=black, postaction={decorate},
                       decoration={markings,mark=at position .55 with {\arrow[draw=black]{<}}}},
    fermionnoarrow/.style={draw=black},
    gluon/.style={decorate, draw=black,decoration={coil,amplitude=4pt, segment length=4pt}},
    scalar/.style={dashed,draw=black,
      postaction={decorate},decoration={markings,mark=at position .55
        with {\arrow[draw=black]{>}}}}, 
    scalarbar/.style={dashed,draw=black,
      postaction={decorate},decoration={markings,mark=at position .55
        with {\arrow[draw=black]{<}}}}, 
    scalarnoarrow/.style={dashed,draw=black},
    electron/.style={draw=black,
      postaction={decorate},decoration={markings,mark=at position .55
        with {\arrow[draw=black]{>}}}}, 
    bigvector/.style={decorate, decoration={snake,amplitude=4pt}, draw},
}

\preprint{MITP/16-137}

\title{Finite remainders of the Konishi at two loops in ${\cal N}=4$ SYM}
\author{Pulak Banerjee$^{a,b}$, Prasanna K.\
  Dhani$^{a,b}$, Maguni Mahakhud$^{b,c}$, V. Ravindran$^{a}$ and Satyajit
  Seth$^{d}$} 

\affiliation{$^a$ The Institute of Mathematical Sciences, Taramani,
  Chennai 600113, India \\ $^{b}$ Homi Bhaba National Institute,
  Training School Complex, Anushakti Nagar, Mumbai 400085, India \\
$^c$ 
Institute of Physics,
P.O: Sainik School,
Bhubaneswar-751005
India \\
$^d$ 
PRISMA Cluster of Excellence, Institut f\"{u}r Physik, Johannes
Gutenberg-Universit\"{a}t Mainz, D\,-\,55099 Mainz, Germany}

\abstract{
We present three point form factors (FF) in ${\cal N}=4$ Super Yang Mills theory for both the half-BPS and 
the Konishi operators at two loop level in the `t Hooft coupling
using Feynman diagrammatic approach.  We have chosen on shell final states consisting of  $g \phi \phi$ and 
$\phi \lambda \lambda$, where $\phi,\lambda, g$ are scalar, Majorana fermion and gauge fields respectively.     
The computation is done both in the modified dimensional reduction as well as in
the four dimensional helicity scheme.     
We have studied the universal structure of infrared (IR) singularities in these FFs using Catani's IR subtraction
operators.  Exploiting the factorisation property of the IR singularities and following BDS like ansatz for
the IR sensitive terms in FFs, we determine the finite remainders of them.   
We find that the finite remainders of FFs of the half-BPS for both the choices of final states give not only identical results but also
contain terms of uniform transcendentality of weight two and four at one and two loop levels, respectively. 
In the case of the Konishi operator, the finite remainders depend on the external states and do not exhibit uniform transcendentality.
However, surprisingly, the leading transcendental terms for $g \phi \phi$ agree with that of the half-BPS.     
We have demonstrated the role of on shell external states for the FFs in the context of maximum transcendentality principle.
}

\begin{document}
\allowdisplaybreaks[4]
\unitlength1cm

\maketitle
\flushbottom


\def\D{{\cal D}}
\def\DD{\overline{\cal D}}
\def\g{\overline{\cal G}}
\def\gm{\gamma}
\def\M{{\cal M}}
\def\ep{\epsilon}
\def\epm1{\frac{1}{\epsilon}}
\def\epm2{\frac{1}{\epsilon^{2}}}
\def\epm3{\frac{1}{\epsilon^{3}}}
\def\epm4{\frac{1}{\epsilon^{4}}}
\def\unM{\hat{\cal M}}
\def\ashat{\hat{a}_{s}}
\def\asmur{a_{s}^{2}(\mu_{R}^{2})}
\def\sigbar{{{\overline {\sigma}}}\left(a_{s}(\mu_{R}^{2}), L\left(\mu_{R}^{2}, m_{H}^{2}\right)\right)}
\def\sigbarn{{{{\overline \sigma}}_{n}\left(a_{s}(\mu_{R}^{2}) L\left(\mu_{R}^{2}, m_{H}^{2}\right)\right)}}
\def\sigh{\hat{\sigma}}
\def\unas{ \left( \frac{\hat{a}_s}{\mu_0^{\epsilon}} S_{\epsilon} \right) }
\def\rnM{{\cal M}}
\def\bt{\beta}
\def\cD{{\cal D}}
\def\cC{{\cal C}}
\def\ca{\text{\tiny C}_\text{\tiny A}}
\def\cf{\text{\tiny C}_\text{\tiny F}}
\def\ct{{\red []}}
\def\sv{\text{SV}}
\def\murOmu{\left( \frac{\mu_{R}^{2}}{\mu^{2}} \right)}
\def\bb{b{\bar{b}}}
\def\bt0{\beta_{0}}
\def\bt1{\beta_{1}}
\def\bt2{\beta_{2}}
\def\bt3{\beta_{3}}
\def\gm0{\gamma_{0}}
\def\gm1{\gamma_{1}}
\def\gm2{\gamma_{2}}
\def\gm3{\gamma_{3}}
\def\l{\left}
\def\r{\right}

\newcommand{\dis}{}
\newcommand{\overbar}[1]{mkern-1.5mu\overline{\mkern-1.5mu#1\mkern-1.5mu}\mkern
1.5mu}

\newcommand{\nn}{\nonumber\\}
\newcommand{\be}{\begin{equation}}
\newcommand{\ee}{\end{equation}}
\newcommand{\bea}{\begin{eqnarray}}
\newcommand{\eea}{\end{eqnarray}}


\section{Introduction}
\label{sec:intro}
Scattering amplitudes are the building blocks of many important observables in high energy physics 
and they exhibit remarkable structures of the underlying quantum field theory (QFT).   
Those computed in the electroweak theory and Quantum Chromodynamics (QCD) play an important role in the phenomenology of collider 
physics. For example, scattering of gluons giving rise to multi jets is important to understand physics at very  
high energies.  With the increase in number of loops and legs, amplitudes are not only 
difficult to calculate but also one obtains very large expressions.
Enormous simplifications can be achieved if one uses on-shell helicity amplitude methods. The formalism 
developed by
Parke and Taylor ~\cite{Parke:1986gb} generated a lot of interest in the study of scattering amplitudes  
with a large number of external gluons. Subsequent developments~\cite{Berends:1987me,
Kosower:1989xy,Bern:1990cu, Bern:1991aq, Bern:1992cz,Cristofano:1992cn,Roland:1992cc,Bern:1993wt,Mahlon:1993fe,Mahlon:1993si, Bern:1994zx}
helped to explore more difficult sets of on-shell amplitudes. 
The BCW/BCFW recursion relations~\cite{Britto:2004ap,Britto:2005fq} made tree level computation  simpler. 

An important feature of any QFT is the existence of symmetries, leading
to conserved  currents. Like QCD, the  ${\cal N} = 4$  Super Yang-Mills (SYM) theory 
is a renormalizable gauge theory in  four dimensional Minkowski space. Apart from having all the symmetries of 
QCD there are two extra features of ${\cal N}=4$ SYM, namely supersymmetry and conformal symmetry
\footnote{In the past, conformal symmetry has been used to understand connections between various perturbative results in gauge theories, see \cite{Kataev:2010tm,Kataev:2010du,Kataev:2013vua}}
that make this theory interesting to study. The on-shell amplitudes evaluated in ${\cal N}=4$ SYM theory
have a simpler analytical structure in comparison to QCD amplitudes. The tree level on-shell amplitudes vanish in ${\cal N}=4$ SYM theory 
when all the external legs have identical  helicities. This is a consequence of supersymmetric (SUSY) Ward 
identities ~\cite{Grisaru:1976vm}.
In ~\cite{Bern:1994zx,Bern:1994cg} the one loop supersymmetric on-shell amplitudes having all external particles as gluons  
were computed and  found to be constructible from the tree level amplitudes. 
%

There is another interesting feature of on-shell amplitudes in ${\cal N}=4$ SYM that deserves some discussion.
The Anti-de-Sitter/conformal field theory (AdS/CFT) conjecture proposed by Maldacena~\cite{Maldacena:1997re} 
establishes one to one correspondence between maximally SYM 
theory in four dimensions and gravity in five-dimensional anti-de Sitter space. Due to this conjecture  
certain quantities in the  strong coupling limit of ${\cal N}=4$ SYM can be related to those in the  weakly coupled gravity.
As a consequence of this duality, quantities computed in a perturbative expansion should add up to 
some simple expression. For such quantities, it may be quite likely that the terms in the perturbative expansion 
are related to one another in some simple way. An important step towards this direction was taken in 
~\cite{Anastasiou:2003kj,Eden:1999kh,Eden:2000mv,
Eden:2000vb}.  In ~\cite{Anastasiou:2003kj}, the planar two loop 
four point on-shell amplitude of ${\cal N}=4$ SYM theory was shown to be related to that of one loop counterpart. 
Such a relation for infrared divergent parts of the amplitude can be understood based on the factorisation properties
of gauge theory amplitudes.  Demanding that this property goes through for finite terms as well,  
an ansatz by Bern, Dixon and Smirnov was put forward for 
$n$ point $m$ loop amplitudes, known as BDS conjecture in the article ~\cite{Bern:2005iz}.
Later on, in \cite{Drummond:2007cf}, the BDS ansatz was shown to be a consequence of the anomalous Ward identity. 
The ansatz was confirmed for the four point amplitudes at three loops ~\cite{Anastasiou:2003kj,Bern:2005iz}  and for the five point 
amplitudes up to two loop order ~\cite{Cachazo:2006tj,Bern:2006vw} in perturbative expansion.
In the seminal paper ~\cite{Alday:2007hr} Alday and Maldacena provided a connection  
between on-shell scattering amplitudes in the strong coupling  and Wilson loops defined in dual coordinate space with
light like segments. Drummond, Korchemsky, and Sokatchev \cite{Drummond:2007aua} demonstrated the equality of results of Wilson loop
with four light-like segments and the one-loop four point maximally helicity violating (MHV) amplitude and in \cite{Brandhuber:2007yx}, the equality was established
between n-sided polygon and one-loop n-point MHV amplitude.  
Later Alday and Maldacena showed that 
their Wilson loop calculation does not agree with BDS ansatz when the number of legs is large ~\cite{Alday:2007he,Maldacena:2010kp,Gao:2013dza}.
It is now well known from evidences that the BDS ansatz requires modifications starting from two loop six leg MHV amplitudes   
\cite{Bartels:2008ce,Bern:2008ap}.

In QFT, the other quantity that has been studied in great detail is the 
form factor (FF) of composite operators.  
The FFs are related to 
amplitudes and are constructed by projecting an off-shell initial state onto on-shell final states. 
The off-shell state is made out of 
a composite operator acting on vacuum, while the final states are those of elementary fields in the theory. 
The FFs, like on-shell amplitudes, contribute to scattering cross sections when one or few of the
external particles become off-shell.  In QCD, the study of FFs in perturbation theory provided 
vital informations on the infrared structure of gauge theories in general.    
In recent times there are several studies~\cite{Brandhuber:2010ad,Gehrmann:2011xn,Boels:2015yna,Brandhuber:2012vm,Brandhuber:2011tv,Bork:2011cj} on the FFs in the context of ${\cal N}=4$ SYM 
which may provide more evidences for the AdS/CFT correspondence.
The most widely studied composite operator in ${\cal N}=4$ SYM is the half-BPS operator. 
This operator is  often referred as protected  because using SUSY algebra~\cite{Dobrev:1985vh,Ferrara:1999zg,Ferrara:1999cw,
Minwalla:1997ka,Rasmussen:2000ii},  their anomalous dimension
can be shown to vanish to all orders in perturbation theory.  Hence the FFs of these composite operators look 
relatively simple.  While the study of FFs in ${\cal N}=4$ SYM has gained momentum over past few years,
the first study along this direction was made long ago by van Neerven ~\cite{vanNeerven:1985ja} where he 
had computed the two loop contributions to two point FF of a half-BPS operator belonging to the stress-energy 
supermultiplet.  The corresponding three loop result was reported in ~\cite{Gehrmann:2011xn}. 
The authors of ~\cite{Gehrmann:2011xn} have shown a remarkable connection between their results and
those in non-supersymmetric $SU(N)$ gauge
theory with $n_f$ fermions when the colour factors satisfy the condition $C_A=C_F=N$ and $n_f T_f=N/2$.  
In particular, upto three loop level, one finds that the results show uniform transcendentality order by order  
in perturbation theory. 

The other type of composite operators that are of interest belongs to non-BPS 
operators. One such representative  is the Konishi operator.
The Konishi  operator~\cite{Konishi:1983hf} is the 
primary operator  of the Konishi  supermultiplet. It is the simplest gauge invariant Wilson operator in ${\cal N}=4$ SYM that 
receives anomalous dimension to all orders in perturbation theory. 
The Konishi supermultiplet has interesting 
connections with string theory. In an $AdS_{5}\,\times S^{5}$ background having type IIB superstring excitations, the first 
string level of the spectrum corresponds to the Konishi supermultiplet ~\cite{Andrianopoli:1998jh}. All the operators of 
the Konishi supermultiplet have same anomalous dimension. The few terms of the anomalous
dimension in the strong coupling region using semiclassical quantisation of short strings in 
$AdS_{5}\,\times S^{5}$ background are available . Similarly, when the coupling is weak, the results
are available to fifth order in perturbation theory.  The  results 
upto two loop were reported in ~\cite{Anselmi:1996mq,Eden:2000mv,Bianchi:2000hn}, third order result was obtained   
in ~\cite{Kotikov:2004er,Eden:2004ua}. At four and five loops, they   
were predicted in ~\cite{Bajnok:2008bm,Bajnok:2009vm} using integrable string sigma model. The four loop result was later confirmed using ${\cal N}=1$ 
Feynman super-graphs method  in ~\cite{Fiamberti:2007rj, Fiamberti:2008sh} and also by Feynman diagrammatic 
approach in ~\cite{Velizhanin:2008jd}. 
Confirmation of anomalous dimension at fifth order was obtained in ~\cite{Eden:2012fe}.  

Over the last few years, there are several results~\cite{Wilhelm:2014qua,Loebbert:2015ova,Brandhuber:2016fni,Loebbert:2016xkw} on  
n-point FFs for the Konishi operator in the weak coupling limit of ${\cal N}=4$ SYM. 
Through them we can, not only confirm the results on the known anomalous dimension but also
study the perturbative infrared structure of off-shell amplitudes in general.  
In addition, they provide testing grounds for various modern techniques that were developed
to compute on-shell amplitudes, namely the on-shell unitarity based methods that use unitarity \cite{Bern:1994zx,Britto:2005fq} and 
various recursion relations \cite{Bern:1994cg,Bern:2007dw}
The results for the
one-loop two point, two loop two point and one-loop three point FFs were presented in ~\cite{Nandan:2014oga}.
In our earlier paper ~\cite{Ahmed:2016vgl}, we have computed the three loop two point FF for the Konishi operator and also
predicted upto  $1/\epsilon$ ($\epsilon=d-4$) pole at four loop order in $d$ dimensions.

In QFT, on-shell amplitudes beyond leading order are plagued with ultraviolet (UV) and infrared (IR)
divergences. The FFs of composite operators in addition acquire 
UV divergences due to the compositeness of the operator. By properly multiplying an
overall operator renormalization factor, we can eliminate these divergences.
The study on the IR structure of on-shell amplitudes and FFs is not only wide but also rich.  
In QCD, the IR structure of the two point FFs 
using gauge invariance, factorization and renormalization group invariances were demonstrated
through the KG-equation ~\cite{Sudakov:1954sw,Mueller:1979ih, Collins:1980ih,Sen:1981sd}.   
Through his pioneering contribution, Catani 
~\cite{Catani:1998bh} 
predicted the universal IR divergences for n-point QCD amplitudes up to two loops.   
Later on  Catani's proposal  of prediction of these divergences was successfully related to 
factorization and  resummation properties of QCD amplitudes in ~\cite{Sterman:2002qn}.  
The generalisation of the results in ~\cite{Catani:1998bh} and ~\cite{Sterman:2002qn} for 
all loop in $SU(N)$ gauge theory having $n_f$ light flavours in terms of cusp, collinear and soft anomalous dimensions
was formulated by Becher and Neubert ~\cite{Becher:2009cu} and independently by Gardi and Magnea ~\cite{Gardi:2009qi}.
These results are not only useful to understand the perturbative structure of multi-loop multi-leg processes in QCD but also
provide deep insight into the structure gauge theories in general.  In addition, iterative structure
observed among the IR sensitive quantities in the perturbative expansion of QCD amplitudes lead to a very successful
program to achieve resummed observables that have played an important role in understanding the physics at
high energies.  In many respects, ${\cal N}=4$ SYM and $SU(N)$ gauge theory with massless fermions show several similarities. 
The fact that there is no coupling constant renormalization in ${\cal N}=4$ makes it more tractable than
$SU(N)$ gauge theory.  Many of the perturbative results in ${\cal N}=4$ can be easily understood
using the well known frameworks in $SU(N)$ gauge theory that deal with organising UV and IR divergences in perturbation
theory.  For example, in \cite{Anastasiou:2003kj,Bern:2005iz}, the IR structure of $SU(N)$ gauge theory along with the explicit results
in ${\cal N}=4$ was readily used to conjecture an all order resummation of finite terms of planar amplitudes.    
For FFs of both the BPS and non-BPS operators in  ${\cal N}=4$ SYM, one
can readily use the KG equation to unravel the IR structure of these form factors in terms of
IR anomalous dimensions as is done for FFs in $SU(N)$ gauge theory.    
In other words, it opens up a rich avenue for investigation of universal structures of FFs in ${\cal N}=4$ SYM and 
gives hints for studying resummation methods.
The FFs of the BPS and the Konishi operators are expressible in terms of cusp, soft and collinear
anomalous dimensions. In our earlier paper  ~\cite{Ahmed:2016vgl} we have studied third order corrections to 
 two point FFs for both the half-BPS and  the Konishi operator.   
The FFs were shown to obey the KG-equation . The universality of IR 
 structure of FF was confirmed as the cusp, soft and collinear divergences obtained to third
 order were found to be same for both of the operators. 

In this paper we will extend our study to three point FFs of the half-BPS and the Konishi operators.
We will present for the first time the full analytical results for the FFs as well as 
finite remainder functions of the Konishi operator up to two loops for both choices of external states $f=g \phi \phi,
\phi \lambda \lambda$. 
In order to show the universality of IR divergences we will use     
Catani's prediction for $SU(N)$ after suitably modifying it to the case of ${\cal N}=4$ SYM and
 demonstrate that Catani's subtraction operators in ${\cal N}=4$ removes all the IR singularities.
This will be a clear demonstration of exponentiation of 
IR singularities of FFs in ${\cal N}=4$ SYM theory, a phenomenon that is well known to occur in QCD ~\cite{Catani:1998bh,
Sterman:2002qn,Becher:2009cu,Gardi:2009qi}.
Following BDS type of analysis, the exponentiation of IR singularities 
in FFs guarantee to yield a finite result the so called finite remainder function. 
A similar study for the amplitudes in ${\cal N}=4$ can be found in ~\cite{Bern:2005iz}.
After subtracting the IR divergent parts,
we present the finite part of the FFs for both the half-BPS and the Konishi operators computed using
the external states $f=g\phi\phi, \phi\lambda\lambda$.  In addition, we will present
their corresponding finite remainder functions.
It is worth mentioning here the work of 
Brandhuber,Travaglini, and Yang 
\cite {Brandhuber:2012vm} who used the generalised unitarity method to build the two loop three point  FFs for 
the half-BPS operator between the states $f=g\phi\phi$. 
Their result for the finite remainder function expressed in terms of loop integrals was then compared numerically against
the one obtained 
using the  method of symbols that uses the
concept of coproduct ~\cite{Goncharov:2010jf}.  
We will reproduce their results for the half-BPS operator at 
two loops for two sets of external states.  Our results are expressed fully in terms of 
Harmonic Polylogarithms (HPL). After quiet a bit of simplifications, we have shown that 
HPLs that appear in our expression are of 
transcendentality four. This is expected as half-BPS operators contain only the highest transcendental terms. 
We will discuss the relation between the highest transcendental terms of finite terms of
FFs of the Konishi operator computed using Catani's as well as BDS approaches and the corresponding ones of the
half-BPS operator.  We will report the method as well as the important findings
for the three point FFs of the half-BPS and the Konishi operators up to two loop level in the subsequent sections of this paper.

The paper is organised as follows.  In section \ref{sec:Lagrangian}, \ref{sec:FF} we briefly discuss the Lagrangian, its field 
content for ${\cal N}=4$ SYM, and the definition of  
form factors of both the half-BPS and the Konishi operators.  In section \ref{sec:methodology}, we present 
the method that we follow to perform the computation of the FFs up to two loop level
in perturbation theory.  Various methods of regularisations,
a discussion on UV and IR structures of the three point FFs and 
the results for their finite parts are presented in section \ref{sec:regularisation}.  
The finite remainder functions are presented in section \ref{sec:FR} after exponentiating the FFs and following BDS approach.
We have devoted  section \ref{sec:LT} to the discussion of  the maximum transcendentality principle in the context of
both the half-BPS and the Konishi FFs.  Finally we conclude in section \ref{sec:conclusion}.  
\section{Theoretical framework}
\label{sec:theory}
\subsection{The ${\cal N}=4$ SYM Lagrangian}
\label{sec:Lagrangian}
In this section we briefly describe the Lagrangian and  its associated fields. 
The Lagrangian for ${\cal N}=4$ SYM in four dimensional Minkowski space reads as ~\cite{Brink:1976bc,Gliozzi:1976qd,
Jones:1977zr,Poggio:1977ma},
\begin{align}
\label{theory1}
\mathcal{L}^{{\cal N} = 4}_{\text{SYM}} = &-\frac{1}{4}G_{\mu\nu}^a G^{\mu\nu a} - \frac{1}{2\xi}(\partial_{\mu}A^{a\mu})^2 + \partial_{\mu}\bar{\eta}^a 
D^{\mu}\eta_a + \frac{i}{2}\bar{\lambda}^a_m\gamma^{\mu}D_{\mu}\lambda^a_m + \frac{1}{2}(D_{\mu}\phi^a_i)^2 
\nonumber\\
&+ \frac{1}{2}(D_{\mu}\chi^a_i)^2 - \frac{g}{2}f^{abc}\bar{\lambda}^a_m[\alpha^i_{m,n}\phi^b_i +
 \gamma_5\beta^i_{m,n}\chi^b_i]\lambda^c_n - \frac{g^2}{4}\Big[(f^{abc}\phi^b_i\phi^c_j)^2 
\nonumber\\
&+ (f^{abc}\chi^b_i\chi^c_j)^2 + 2 (f^{abc}\phi^b_i\chi^c_j)^2\Big].
\end{align}
In the  above Lagrangian all the fields transform in adjoint representation and hence carry  SU(N) color indices $a,b,c$.
 $A^{a,\mu}$,  $\eta^a$  represent the gauge and ghost fields respectively while $\xi$ is the gauge fixing parameter.
The Majorana fields are denoted by $\lambda^a_m$, with $m=1,...,4$ denoting their generation type.
The scalar and pseudoscalar fields $ \phi^a_i$ and $\chi^a_i $, which help to 
maintain fermionic and bosonic degrees of freedom same,  interact identically with 
the gauge fields. The indices  i,\,j represent different types of scalars and  pseudoscalars in theory.
In four dimensions, $i,j=1,2,3$.
The gluonic field strength is given by 
\begin{equation*}
\label{theory2}
G_{\mu\nu}^a = \partial_{\mu}A_{\nu}^a - \partial_{\nu}A_{\mu}^a + g f^{abc}A_{\mu}^b A_{\nu}^c,
\end{equation*}
while  covariant derivative  is  $D_{\mu} = \partial_{\mu} - i g T^a A_{\mu}^a$. $g$ is the coupling constant.
The matrices $T^a$ satisfy  
\begin{equation*}
\label{theory3}
[T^a,T^b]_{-} = i f^{abc}T^c,
\end{equation*}
where $f^{abc}$ is the totally antisymmetric structure constant of the group algebra. The generators are normalized as 
$\text{Tr}(T^a\,T^b) = 1/2\,\delta^{a\,b}$. The six antisymmetric matrices $\alpha$ and $\beta$
satisfy the following commutation and anticommutation relations
\begin{equation}
\label{theory4}
[\alpha^i, \alpha^j]_{+} = [\beta^i, \beta^j]_{+} = -2 \delta^{ij} \mathbb{I},\quad  [\alpha^i, \beta^j]_{-} = 0.
\end{equation}

The traces of the above mentioned  matrices, which we shall require in our computation  are given by
\begin{equation}
\label{theory5}
\text{tr}(\alpha^i) = \text{tr}(\beta^i) = \text{tr}(\alpha^i\beta^j)= 0,\quad \text{tr}(\alpha^i\alpha^j) = \text{tr}(\beta^i\beta^j) = -4 \delta^{ij}.
\end{equation}
The UV and IR divergences arising during our computation need to be properly regularized. We will be following the supersymmetry preserving
dimensional reduction scheme ($\overline{DR}$) ~\cite{Siegel:1979wq,Capper:1979ns}.
Thus the following relations are known to  satisfy in $d=4+\epsilon$
\begin{equation}
\label{theory6}
\alpha^i\alpha^i = \beta^i\beta^i = \Big(-3 +  \frac{\epsilon}{2}\Big) \mathbb{I} \,\,, 
\alpha^i\alpha^j\alpha^i  = \alpha^j \Big(1- \frac{\epsilon}{2}\Big) \mathbb{I} \,\,,
\quad \beta^i\beta^j\beta^i  = \beta^j \Big(1- \frac{\epsilon}{2} \Big)  \mathbb{I}.
\end{equation}
\subsection{Form Factors}
\label{sec:FF}
Composite operators provide valuable information about the underlying QFT through their
correlation functions and form factors.  
Those in the ${\cal N}=4$ SYM are special in the context of AdS/CFT correspondence.   As we have
discussed in the introduction, the BPS and non-BPS composite operators have attracted the attention
for a long time.  In the following, we will study structure of the three point form factors of each operator,
namely one half-BPS and the Konishi belonging to non-BPS type up to two loop level in perturbation theory.
Thanks to supersymmetry the half-BPS one is protected, hence it is UV finite.  In other words, it does not
require any overall UV renormalisation constant to all orders in perturbation theory.  On the other hand,
the Konishi operator develops UV divergences which can be renormalised away by over all renormalisation constant.
Consequently, unlike the half-BPS, the Konishi operator develops anomalous dimension that can be computable in powers of
the coupling constant $g$.  Let us denote these operators by ${\cal O}^\rho$ with $\rho = BPS,{\cal K}$ respectively.
The half-BPS operator that we will study is given by ~\cite{Bergshoeff:1980is,vanNeerven:1985ja}
\begin{equation}
\label{Op1}
{\cal O}^{\rm{BPS}}_{rt} = \phi^a_r\phi^a_t - \frac{1}{3}\delta_{rt}\phi^a_s\phi^a_s,
\end{equation}
 where the symbol $a \in [1,N^2-1]$ represents the $SU(N)$ adjoint colour index, while $\phi_r$ are the scalar fields.
The factor 1/3 has been used to maintain tracelessness property in 4 dimension.  
The Konishi operator ($\rho = {\cal K}$) which is   
the primary  operator of the Konishi supermultiplet  is given by
 \begin{equation}
 \label{Op2}
{\cal O}^{\cal K} = \phi^a_r\phi^a_r + \chi^a_r\chi^a_r .
 \end{equation} 
The fields $\phi$ and $\chi$ represent the scalar and pseudoscalar types respectively, with $r \in [1,3]$. It is one of the non-BPS operators 
that has been studied extensively. An interesting feature of this operator could 
be noted in context of operator product expansion.
During  loop calculations the leading asymptotic behaviour of the four point correlation 
function at short distances was  found to be regulated by the Konishi operator.  

The interaction parts of the Lagrangian that describe the couplings of external currents $J$ ($J^{\text{BPS}}_{rt}$,  $J^{{\cal K}}$) 
to the operators \eqref{Op1}, \eqref{Op2} 
are given in 
\begin{equation}
\label{Op3}
\mathcal{L}^{\text{BPS}}_{\text{int}} = J^{\text{BPS}}_{rt}O^{\text{BPS}}_{rt}, \quad  \mathcal{L}^{{\cal K}}_{\text{int}} 
= J^{{\cal K}}O^{{\cal K}}.
\end{equation}

We are interested to study three point FFs of the above two operators for two specific final states, $f = { g \phi \phi}$ and $f = {\phi \lambda \lambda}$.
We define the renormalized FF of composite operators ${\cal O}^{\rho}$ between the set of states, $f$ by
\begin{equation}
\label{Op4}
  \mathcal{F}_{f}^{\rho, (n)} \equiv \frac{\langle \mathcal{M}_{f}^{\rho,(0)}|  \mathcal{M}_{f}^{\rho,(n)}\rangle}
  {\langle \mathcal{M}_{f}^{\rho,(0)}| \mathcal{M}_{f}^{\rho,(0)}\rangle}
\end{equation}
where $n = 0,1,2, ...$ and  $\rho = $  $\rm{BPS}$ or ${\cal K}$. 
In the above expression $ | \mathcal{M}_{ f }^{\rho,(n)}\rangle$ is the $O({a}^n)$ contribution to the transition
amplitude for an off-shell state created by ${\cal O} ^{\rho}$ to the three on-shell states denoted by 
$f = { g \phi \phi}, {\phi \lambda \lambda}$. 
Here $a$ is the `t Hooft coupling \cite{Bern:2005iz} given by
\begin{equation}
\label{eq:tHcoup}
{a} = \frac{g^2 N}{(4 \pi)^2}\left(4 \pi e^{-\gamma_E}\right)^{-{\epsilon \over 2}}.
\end{equation}
where $\gamma_E=0.5772\cdot\cdot\cdot$, is the Euler constant.
Since we work in momentum space, these amplitudes are
dependent on momentum $q$ of the off-shell state and momenta $p_i,i=1,2,3$ of the on-shell final states. 
The corresponding Mandelstam variables are given by
 \begin{equation}
  s\equiv(p_1+p_2)^2,\quad t\equiv(p_2+p_3)^2,\quad u\equiv(p_1+p_3)^2
  \end{equation}
  which obey the relation
  \begin{equation}
  s+t+u = q^2=Q^2.
  \end{equation}
 We also define the following dimensionless invariants, which we will use to describe our
 results in terms of  HPL ~\cite{Remiddi:1999ew}
   and 2dHPL ~\cite{Gehrmann:2000zt,Gehrmann:2001ck} as
  \begin{equation}
  x\equiv\frac{s}{Q^2},\quad y\equiv \frac{u}{Q^2}, \quad z\equiv \frac{t}{Q^2},
  \end{equation}
 where $(x,y,z)$ lie between $0 $ and $1$ and satisfy the condition
  \begin{equation}
  x+y+z = 1.
 \end{equation}
In the following, we will describe how we have performed our computation of the FFs of these two
operators between various sets of states $f$ at two loop level.

\subsection{Methodology of calculation}
\label{sec:methodology}
In this section we describe the methodology to compute the FFs, as defined in eq. \eqref{Op4}, for the three external
on-shell states up to second order in perturbative expansion. 
There has been a vast amount of developments 
over last two decades to compute scattering amplitudes, form factors etc to very high orders in perturbative QCD.   
In the present context, it is worth mentioning earlier works on the two loop three point FF computations 
for the scattering cross section  $e^{+}\, e^{-} \rightarrow 3\, \text{jets}$ ~\cite{Garland:2002ak},
Higgs decays $H \rightarrow g+g+g$ and 
$H \rightarrow g+q+\bar{q}$ in  ~\cite{Gehrmann:2011aa,Ahmed:2014pka}.
 Decays of a massive spin-2 resonance to 3 gluons as well as fermion anti-fermion
 and a gluon can be found in ~\cite{Ahmed:2014gla,Ahmed:2016yox} and some other works~\cite{Ahmed:2016qhu,Ahmed:2016qjf}. 
 Computation of the production of massive spin 1 bosons at two loop level in QCD can be found in the literature 
 \cite{Gehrmann:2013vga}. 
Similar calculation
 can be found in \cite{Ahmed:future}.  These computations often use modern helicity methods 
to compute the amplitudes first and then 
the resulting large number of two loop integrals are reduced to fewer master integrals using 
an elegant integral reduction technique, called
integration-by-parts, pioneered
by Tkachov and Chetyrkin in ~\cite{Tkachov:1981wb,Chetyrkin:1981qh} supplemented by    
identities that are obtained using Lorentz invariance  ~\cite{Gehrmann:1999as}. 
Dedicated efforts were made in order to compute these two loop master integrals
thanks to the method of differential equations that these master integrals satisfy.
Thanks to various modern techniques that followed, such as
the `on-shell'
techniques, where one uses recursion relations ~\cite{Britto:2004ap,Britto:2005fq}
 and unitarity  ~\cite{Bern:1994zx,Bern:1994cg} relations, the loop computations 
with many legs become tractable.  In the context of ${\cal N}=4$ SYM,  the loop induced 
contributions to 
$n$-point FF for the half-BPS operator were presented in ~\cite{Brandhuber:2010ad,Bork:2010wf},
also in \cite{Brandhuber:2011tv,Bork:2011cj}. For the half-BPS operator made up of $k$ scalar
 fields,  the results for FFs can be found in \cite{Bork:2010wf,Penante:2014sza,Brandhuber:2014ica}.
 The n-point MHV
 as well as next to MHV FFs, up to first order in perturbative expansion are presented in
 \cite{Brandhuber:2010ad,Brandhuber:2011tv,Bork:2011cj,Bork:2012tt,Engelund:2012re,Bork:2014eqa}. 
 At strong coupling limit, FFs have also been studied via AdS/CFT correspondence in ~\cite{Alday:2007he}.
We will follow the standard Feynman diagrammatic approach to compute three point FFs at two loop level.
We perform the computation in $d=4+\epsilon$ dimensions so that divergences appearing in 
the intermediate stages can be regularised.  We will discuss details of the reguralization
schemes and the nature of various singularities that appear in the computation.

 We have used QGRAF ~\cite{Nogueira:1991ex} to generate Feynman diagrams for the computation of  
$ | \mathcal{M}_{ f }^{\rho,(n)}\rangle$ where $\rho={\rm BPS},{\cal K}$,  $f$ is one of the sets
 $f=\phi\lambda\lambda$ and $f=g\phi\phi$.
For the BPS operator at two loop, there are 1024 diagrams when $f=g\phi\phi$ and 616 diagrams for $f=\phi\lambda\lambda$.
At same order in perturbative expansion, for the Konishi operator with $f=g\phi\phi$, we have 1220 diagrams
while for $f=\phi\lambda\lambda$, we get 825 diagrams.
While dealing with the Majorana fermions care must be taken to ensure its self conjugacy property. Being its own antiparticle,
we have to modify QGRAF output in such a way that the fermion flows remain intact. We  use in-house PYTHON code to
deal with the issue. The raw output from QGRAF is converted to a suitable format for further computations. The 
simplification of matrix elements involving the Lorentz, Dirac, colour and generation indices is done by employing
a set of in-house routines based on the symbolic manipulating program FORM ~\cite{Vermaseren:2000nd}.
For external on-shell gluon legs the sum over polarizations is carried out using
\begin{equation}
\sum_s \epsilon^{\mu}(p_1; s) \epsilon^{*\nu}(p_1;s) = -g^{\mu\nu} + \frac{p_1^\mu q^\nu + p_1^\nu q^\mu }{p_1.q}
\end{equation}
where $q^\mu $ is an arbitrary light-like four vector. 
These matrix elements comprises of huge number of one and two loop scalar Feynman integrals. 
We have classified them so that they  belong to a few integral sets. Using Reduze \cite{vonManteuffel:2012np}, the loop 
momenta in various integrals can be shifted suitably in order to belong to these sets.  
All the one loop diagrams are reduced to three sets 
expressed in a condensed way as follows:
\begin{equation}
\{ \mathcal P, \, \mathcal {P}_{i}, \, \mathcal P_{i,i+1},\, \mathcal P_{i,i+1,i+2} \}.
\end{equation}
Here  for each set, $i$ takes any one value from $ \{1,2,3\}$ whose elements are arranged cyclically. 
$\cal {P}$'s are
\begin{equation}
\mathcal {P} =k_1^2,\, \mathcal {P}_i=(k_1 - p_i)^2,\, \mathcal {P}_{i,j}=(k_1 - p_i - p_j)^2,\, 
\mathcal {P}_{i,j,k}=(k_1 - p_i - p_j  - p_k)^2.
\end{equation}
This becomes more complicated at two loops.
At two loops, we have nine independent Lorentz invariants constructed 
out of two loop momentas $k_1$ and $k _2$, namely $\{ k_ \delta \, . k_ \gamma\,, k_ \delta\, . p_j \}$,
$\delta, \gamma = 1,2; j = 1,2,3$. After shifting of loop momenta we can fit all the Feynman integrals
into six sets. The first three can be written in a condensed notation as:
\begin{equation}
\{ \mathcal P _ 0,\, \mathcal P _ 1,\, \mathcal P _ 2,\, \mathcal P _ {1:i},\, \mathcal P _ {2:i},\,  
\mathcal P _ {1:i,i+1},\, \mathcal P _ {2:i,i+1},\, \mathcal P _ {1:i,i+1,i+2},\, \mathcal P _ {2:i,i+1,i+2} \},
\end{equation}
and  the three remaining sets can be expressed as follows:
\begin{equation}
\{ \mathcal P _ 0,\, \mathcal P _ 1,\, \mathcal P _ 2,\, \mathcal P _ {1:i},\, \mathcal P _ {2:i},\,  
\mathcal P _ {0:i+2},\, \mathcal P _ {1:i,i+1},\, \mathcal P _ {2:i,i+1},\, \mathcal P _ {1:i,i+1,i+2} \},
\end{equation}
where for each set, $i$ takes any one value from $ \{1,2,3\}$ whose elements are arranged 
cyclically. The elements of the six sets are
\begin{align}
&\mathcal P_0 =(k_1 - k_2)^2, \qquad \qquad \,\,\,  \mathcal P_{\beta} = k_\beta^ 2, 
\nn&
\mathcal P_{\beta:i} = (k_\beta - p_i)^2, \qquad \qquad \mathcal P_{\beta:i,j} =(k_\beta -p_i -p_j)^2, 
\nn &
\mathcal P _{0:i} = (k_1 - k_2 - p_i)^2, \qquad  \mathcal P_{\beta:i,j,k} = (k_\beta -p_i -p_j -p_k)^2. 
\end{align}
Although the number of scalar integrals are huge in number, most of them 
are however,  related to one another. Using integration-by-parts  ~\cite{Tkachov:1981wb,Chetyrkin:1981qh} and 
Lorentz invariance identities ~\cite{Gehrmann:1999as} they can be reduced to a smaller set of topologically different integrals, 
popularly  known as master integrals (MI). We have used a Mathematica based
package LiteRed ~\cite{Lee:2008tj,Lee:2012cn} to uncover the set of MIs. For one and two loop the number of MIs are 7 and
89 respectively. These MIs were computed by Gehrmann and Remiddi analytically as expansion in 
Laurent series in $\epsilon$ in the references
~\cite{Gehrmann:2000zt,Gehrmann:2001ck}.
%
%
\section{Regularization of  singularities}
\label{sec:regularisation}
${\cal N}=4$ SYM is ultraviolet finite theory in four dimensions, which guarantee the UV finiteness
of quantities like on shell amplitudes, FFs of protected operators and correlation functions of 
elementary fields etc.  Unprotected operators develop UV divergences.  
Due to the presence of massless fields
in the theory, on shell amplitudes and both protected as well as unprotected FFs 
are sensitive to infrared divergences order by order in
perturbation theory.    

In QFT, the method of dimensional regularization (DR) has been quite successful
in regulating both UV as well as IR singularities.  However, different variants or schemes of DR exist in the
literature.
In the scheme proposed by 't Hooft and Veltman ~\cite{'tHooft:1972fi}, called DR scheme,
the unphysical gauge bosons are treated in $4+\epsilon$ dimensions with  $2+\epsilon$ helicity states 
but the physical ones in 4 dimensions having 2 helicity states.  
The conventional dimensional regularization (CDR) scheme proposed by Ellis and Sexton ~\cite{Ellis:1985er}
treats both the physical and unphysical gauge fields in  $4+\epsilon$ dimensions.
The four dimensional helicity (FDH) scheme advocated in ~\cite{Bern:1991aq,Bern:2002zk} assumes both 
the physical and unphysical states in 4 dimension.
The common among all these schemes is that the loop integrals are performed in 
$4+\epsilon$ dimensions.  Of course, all these schemes can be related to each other using suitable
finite renormalisation constants. 
FDH scheme has been the most popular one in supersymmetric theories.   
It was observed in ~\cite{Nandan:2014oga} that for certain dimension dependent operators, extra care
is needed to get the correct result.  One such operator is the Konishi operator.
In ~\cite{Nandan:2014oga}, the authors propose a prescription to correct the result for the
Konishi operator computed in 
FDH scheme.  In order to obtain the correct result for its FF, their prescription suggests that
one has to take the difference of FFs of the Konishi and the half-BPS operators, multiply
the ratio $\Delta^{BPS}_{{\cal K}} = n_{g,\epsilon}/2n_g$  and add to the FF of the half-BPS, that is
\begin{eqnarray}
{\cal F}^{\cal K}_f = {\cal F}^{BPS}_f + \Delta^{BPS}_{\cal K} ({\cal F}^{\cal K}_f - {\cal F}^{BPS}_f)
\end{eqnarray}
where $n_g=3$ is the number of scalar or pseudo scalar fields and $n_{g,\epsilon} = 2 n_g - \epsilon$.
It is shown that this prescription correctly reproduces the anomalous dimension of the Konishi operator up to two loop
level.  There exists another elegant scheme   
called $\overline{DR}$ scheme which we will use in our computation.  
In this scheme, one adjusts the number of generations of scalar and pseudo scalar fields such that 
the resulting bosonic degrees of freedom adds to that of fermions in order to preserve supersymmetry.
Since the gauge fields have $2+\epsilon$ degrees of freedom, there should be $n_g-\epsilon/2$ scalars
and $n_g-\epsilon/2$ pseudoscalar in the regularised version of the theory so that the total 
number of bosonic degrees of freedom in $4 +\epsilon$ is same as the four dimensional one, namely $2 (1+n_g)$.
The advantage of this scheme over others
is that it can be used even for operators that depend on space time dimensions.
In this scheme, for ${\cal N}=4$ SYM, the number of 
scalar and pseudoscalar fields become $6 - \epsilon$ as opposed to 6. 
Here, gluons have $2+\epsilon$ degrees of freedom while fermions degrees of
freedom remains 8.  
This prescription is consistent with ${\cal N}=4$ supersymmetry since the 
total number of scalar, pseudoscalar and gauge degrees of freedom continues to remain 8.  
In this scheme, all the  traces of $\alpha$,
$\beta$ and Dirac matrices  are done in $4+\epsilon$ dimensions.  
In ~\cite{Ahmed:2016vgl} three loop FFs of the Konishi was computed in the
FDH scheme modified according to ~\cite{Nandan:2014oga}, as well as in the $\overline{DR}$ scheme up to
three loop level in perturbation theory and the results agreed validating the prescription
advocated in ~\cite{Nandan:2014oga}.
In the following we will present the results for the three point FFs of the BPS and the Konishi operators.  
These results were computed both in the ${\overline{DR}}$
as well as FDH schemes.  For the BPS, both the schemes give same results, but for the Konishi, they agree only 
when the results computed in the FDH scheme are modified according to the prescription given in ~\cite{Nandan:2014oga}.

\subsection{UV divergences}
\label{sec:UV}
Since, ${\cal N}=4$ SYM is UV finite, the coupling constant does not get any renormalisation, i.e.,
the beta function of the coupling,  $\beta(a)$ ~\cite{Jones:1977zr,Poggio:1977ma} is identically zero
to all orders in perturbation theory. 
Hence, based on dimensional argument, the coupling constant $\hat a$ that appears in the regularised
Lagrangian in $4+\epsilon$ dimensions is
written in terms of dimensionless coupling $a$ (see eq. (\ref{eq:tHcoup})) through 
\begin{equation}
\label{eq:aren}
  \hat{a} = a \Big(\frac{\mu_0^2}{\mu^2}\Big)^{\frac{\epsilon}{2}},
\end{equation}
The UV divergences arising due to short distance effects  in the composite operators
can by removed by an overall operator renormalization constant $Z^{\rho}\,(a,\mu,\epsilon)$
and hence the  operator's scaling dimension is modified  by  $\gamma^{\rho}$, the anomalous
dimension.
The relation between  $Z^{\rho}\,(a,\mu,\epsilon)$ and  $\gamma^{\rho}$ is 
encapsulated in the following differential equation
\begin{align}
\label{Reg1}
&\frac{d}{d\ln\mu^2} \ln Z^{\rho} =
  \gamma^{\rho} = \sum\limits_{i=1}^{\infty} a^{i}
  \gamma^{\rho}_{i}\,.
\end{align}
Using eq. (\ref{eq:aren}), the solution to the above equation is:
\begin{equation}
\label{Reg2}
Z^{\rho}  = \exp \Big(\sum_{n=1}^{\infty}a^n\frac{2\gamma^\rho_n}{n\epsilon}\Big).
\end{equation}
In our earlier paper ~\cite{Ahmed:2016vgl}, we extracted the $\gamma^{\rho}$ up to three loop order 
using two point three loop form factor of the Konishi and found agreement with the known results.
The anomalous dimension up to 2-loop are in agreement with the existing ones~\cite{Anselmi:1996mq, Eden:2000mv, Bianchi:2000hn}.
We used analog of KG that QCD form factors satisfy  
 ~\cite{Ravindran:2005vv,Bern:2005iz,Ravindran:2006cg} to determine the anomalous dimension $\gamma^{\rho}$.
 The values up to two loop are
\begin{align}
\label{Reg3}
\gamma^{\rm BPS}_{j}=0\,,
\quad \gamma^{\cal K}_{1}= -6\,,\quad  \gamma^{\cal K}_{2} = 24\,.
 \end{align}
The operator renormalised transition amplitudes $|{\mathcal{M}}^{\rho}_f\rangle$ can be expressed in terms of 
UV unrenormalized ones $|{\hat{\mathcal{M}}}^{\rho,(n)}_f\rangle$, expanded in coupling constant $a$ as
\begin{equation}
\label{Reg4}
  |{\mathcal{M}}^{\rho}_{f}\rangle =  Z^{\rho}\,(a,\mu,\epsilon) \Big(\frac{a}{\mu^{\epsilon}}\Big)^{\frac{1}{2}}\Bigg
  (|\hat{\mathcal{M}}^{\rho,(0)}_{f} \rangle+  \Big(\frac{a}     
  {\mu^{\epsilon}}\Big) |\hat{\mathcal{M}}^{\rho,(1)}_{f} \rangle + \Big(\frac{a}     
  {\mu^{\epsilon}}\Big)^2 |\hat{\mathcal{M}}^{\rho,(2)}_{f} \rangle \Bigg).
\end{equation}
For $\rho = \rm{BPS} $, $Z^{BPS}\,(a,\mu,\epsilon) =1$. 
On the other hand, for the Konishi operator, we can express  $|{\mathcal{M}}^{\cal K}_{f}\rangle$ in terms of UV finite matrix elements
$|{\mathcal{M}}^{{\cal K},(j)}_{f} \rangle$ as follows
\begin{equation}
\label{Reg5}
 |{\mathcal{M}}^{\cal K}_{f}\rangle = a^{1/2}  |{\mathcal{M}}^{{\cal K},(0)}_{f}\rangle + a^{3/2}  |{\mathcal{M}}^{{\cal K},(1)}_{f}\rangle
 + a^{5/2} |{\mathcal{M}}^{{\cal K},(2)}_{f}\rangle\,,
\end{equation}
where
\begin{align}
\label{Reg6}
 |{\mathcal{M}}^{{\cal K},(0)}_{f}\rangle &=  \Big(\frac{1}{\mu^{\epsilon}}\Big)^{\frac{1}{2}}  |{\hat{\mathcal{M}}}^{{\cal K},(0)}_{f}\rangle, \nn
 |{\mathcal{M}}^{{\cal K},(1)}_{f}\rangle &=  \Big(\frac{1}{\mu^{\epsilon}}\Big)^ {\frac{3}{2}}\,\Big[ \mu^{\epsilon}\ r_1
  |{\hat{\mathcal{M}}}^{{\cal K},(0)}_{f}\rangle +   |{\hat{\mathcal{M}}}^{{\cal K},(1)}_{f}\rangle\Big], \nn
 |{\mathcal{M}}^{{\cal K},(1)}_{f}\rangle &=  \Big(\frac{1}{\mu^{\epsilon}}\Big)^ {\frac{5}{2}} \Big[ \mu^{2 \epsilon} \Big(\frac{r_1^2}{2}   + r_2 \Big)
 |{\hat{\mathcal{M}}}^{{\cal K},(0)}_{f}\rangle + \mu^{\epsilon} \, r_1 |{\hat{\mathcal{M}}}^{{\cal K},(1)}_{f}\rangle +
 \, |{\hat{\mathcal{M}}}^{{\cal K},(2)}_{f}\rangle \Big],
 \end{align}

with   
\begin{equation}
\label{Reg7}
r_1 = \frac{1}{\epsilon}\, 2\gamma_1^{\cal K},  \qquad r_2 =  \frac{1}{\epsilon}\gamma_2^{\cal K} .
\end{equation}
Substituting the above equations in eq. (\ref{Op4}) we obtain the renormalised FFs for the BPS and
the Konishi operator.  We will present the results in the following sections after a systematic subtraction of 
IR divergences.

\subsection{Universality of IR singularities}
\label{sec:IR}
In gauge theories, the UV renormalized amplitudes and FFs suffer from 
IR divergences due to the presence of massless initial and final states. 
These divergences are of two types, namely soft and collinear. Soft ones arise when the momentum of a 
massless gauge particle in loop has vanishing components. The collinear 
divergences emerge when any massless particle in the loop becomes parallel to the external particle. 
However these amplitudes and FFs are not the observables, instead the cross sections or decay rates made
out of these amplitudes and FFs are the 
observables.  Finiteness of the observables has been proved through Kinoshita-Lee-Nauenberg (KLN) theorem 
~\cite{Kinoshita:1962ur,Lee:1964is}. The IR structure of amplitudes and FFs in QCD is well understood and
in particular, the universal structure of IR divergences was predicted in a seminal paper by 
Catani  ~\cite{Catani:1998bh}
for n-point two loop QCD amplitudes. 
He exploited the iterative structure of singular part of the renormalised amplitudes in QCD
to predict the subtraction operators that capture the universal IR divergences of these amplitudes.
Later on Sterman and Tejeda-Yeomans successfully related Catani's predictions to the factorization
and resummation properties of QCD amplitudes ~\cite{Sterman:2002qn}.
Note that iterative structure is the result of factorisation.  
The generalization of Catani's
proposal for arbitrary number of loops and legs for $SU(N)$ gauge theory
 using soft collinear effective theory was given by Becher and Neubert 
~\cite{Becher:2009cu}.  A similar result was also 
presented by Gardi and Magnea ~\cite{Gardi:2009qi} by analyzing the Wilson lines for hard partons.


Following Catani's proposal, we can express
the UV renormalized amplitudes $ |\mathcal{M}^{\rho,(n)}_{f} \rangle$  
for the three point function up to two loop order in terms of universal subtraction operators  
$\textbf{I}_{f}^{(n)}(\epsilon)$ (see similar analysis for MHV amplitudes in ~\cite{Naculich:2008ys}),  
 \begin{align}
 \label{Reg8}
 &|\mathcal{M}^{\rho,(1)}_{f} \rangle = 2 \,\textbf{I}_{f}^{(1)}(\epsilon) |\mathcal{M}^{\rho,(0)}_{f} \rangle + 
 |\mathcal{M}^{\rho,(1) \text{fin}}_{f} \rangle \, ,
 \nonumber\\
 &|\mathcal{M}^{\rho,(2)}_{f} \rangle  = 2 \,\textbf{I}_{f}^{(1)}(\epsilon) |\mathcal{M}^{\rho,(1)}_{f} \rangle + 
 4 \,\textbf{I}_{f}^{(2)}(\epsilon) |\mathcal{M}^{\rho,(0)}_{f} \rangle + |\mathcal{M}^{\rho,(2)\text{fin}}_{f} \rangle.
 \end{align}
%

In context of  ${\cal N}=4$ SYM theory where all fields are in adjoint representation and $\beta$-function is zero, 
the universal subtraction operators are found to be 
 \begin{align}
 \label{Reg9}
 &\textbf{I}_{f}^{(1)}(\epsilon) = -\frac{1}{2}\frac{e^{-\frac{\epsilon}{2}\gamma_E}}{\Gamma(1+\frac{\epsilon}{2})} 
 \mathcal{V}^{sing}(\epsilon)\Bigg\{\Big(-\frac{s}{\mu^2}\Big)^{\frac{\epsilon}{2}} + \Big(-\frac{t}{\mu^2}\Big)^{\frac{\epsilon}{2}} +
  \Big(-\frac{u}{\mu^2}\Big)^{\frac{\epsilon}{2}}\Bigg\} \, ,
 \nonumber\\
 &\textbf{I}_{f}^{(2)}(\epsilon) =-\frac{1}{2} \Big(\textbf{I}_{f}^{(1)}(\epsilon)\Big)^2 + 
 K\frac{e^{\frac{\epsilon}{2}\gamma_E}\Gamma(1+\epsilon)}{\Gamma(1+\frac{\epsilon}{2})} \textbf{I}_{f}^{(1)}(2\epsilon) 
 + \textbf{H}^{(2)}_{f}(\epsilon),
 \end{align}
 where $\mathcal{V}^{sing}(\epsilon) = \frac{A_{1}}{\epsilon^2}$ results from the dipole
structure of the infrared singularities. The constant $A_1$ is the first coefficient of the cusp anomalous dimension and found to be 4 in $\mathcal{N}=4$ SYM.  Note that the constant $K$ and 
$\textbf{H}^{(2)}_f(\epsilon)$ are unknown at the moment and they can be determined using the
explicit results for $|\mathcal{M}^{\rho,(i)}_{f} \rangle$ computed in this paper demanding the finiteness of 
$|\mathcal{M}^{\rho,(i)\text{fin}}_{f} \rangle$ for $i= 1,2$.

%

Using these subtraction operators, the UV renormalized FF as defined in eq. \eqref{Op4} for $ n = 1,2 $ can be written as 
%
%
%
 \begin{align}
 \label{Reg11}
 &\mathcal{F}_{f}^{\rho, (1)} = 2 \,\textbf{I}_{f}^{(1)}(\epsilon) \mathcal{F}_{f}^{\rho, (0)} + 
 \mathcal{F}_{f}^{{\rho, (1)}, \text{fin}}, 
 \nonumber\\
 &\mathcal{F}_{f}^{\rho, (2)}  = 2 \,\textbf{I}_{f}^{(1)}(\epsilon) \mathcal{F}_{f}^{\rho, (1)} + 
 4 \,\textbf{I}_{f}^{(2)}(\epsilon) \mathcal{F}_{f}^{\rho, (0)} + \mathcal{F}_{f}^{{\rho, (2)}, \text{fin}} .
 \end{align}
Substituting the UV finite $ \mathcal{F}_{f}^{\rho, (i)}$ in eq. \eqref{Reg11} 
and demanding finiteness of 
$ \mathcal{F}_{f}^{{\rho, (i)}, \text{fin}}$ 
we can determine $K$ and $\textbf{H}^{(2)}_{f}(\epsilon)$.
These constants $K$ and $\textbf{H}^{(2)}_{f}(\epsilon)$ 
obtained independently from the results of FFs of the BPS as well as from the Konishi operators,  
are found to be independent of the type of operator.
They are given by
\begin{eqnarray}
\label{eq:KH}
K = - \zeta_2, \quad
\textbf{H}^{(2)}_{f}(\epsilon) = -\frac{3}{4\epsilon}  \zeta_3.
\end{eqnarray}
The factor $3$ in $\textbf{H}^{(2)}_f(\epsilon)$ is due to the presence of three massless particles in the final state $f$ all giving same
contribution.  Like in QCD, we expect that the universality of these constants to hold for two loop FFs of any composite operator. 
In the appendix, we present remaining finite terms  
$\mathcal{F}_{f}^{{\rho, (n)}, \text{fin}}$ for $n = 1,2$ by setting $ \mu^2=-Q^2 $ in eq. (\ref{Reg11}).  

In the next section we shall discuss about the finite remainders and examine whether it can be written in terms of 
 some simple structure.

\section{Finite Remainders}
\label{sec:FR}
We have seen in the previous section the 
iterative structure of IR divergences for the FFs in terms of Catani's subtraction operators
up to two loop level in perturbation theory.  This is a result of factorisation property of the FFs at long distances.  This is
very similar to on-shell amplitudes in gauge theories.  The factorisation and observed 
universality of these IR divergences can lead to exponentiation of these divergences whose 
coefficients are controlled by certain universal IR anomalous dimensions.  In the case of amplitudes,
we have seen in the context of the BDS ansatz ~\cite{Anastasiou:2003kj,Bern:2005iz}, both the IR and 
finite parts can be exponentiated
and the exponents are expressed in terms of their one loop counter parts multiplied by 
a function that depend on the IR anomalous dimensions such as cusp and collinear ones.  
For amplitudes, the iterative structure of the finite terms breaks down
from two loop six point amplitudes onwards and the deviation is called  
the finite remainder function.   
 
In the case of FFs, the IR divergences exponentiate, thanks to their iterative structure but the finite terms. 
Since, the beta function vanishes to all orders in perturbation theory,
the divergent part of the exponent can be normalised in terms of the corresponding one loop FFs
by properly adjusting the $\epsilon$ at every order.
The finite terms of the FFs do not show any iterative structure,
hence the corresponding finite reminder functions are always non-zero.
For the two point half-BPS operator, the authors of ~\cite{Brandhuber:2010ad} have obtained the finite
reminder function up to two loops.
Later it was extended to cases with more than two external states   
~\cite{Brandhuber:2014ica, Brandhuber:2012vm}.
Following BDS structure, the finite reminder function at two loops is defined by 
\begin{equation}
  \label{FR2}
  \mathcal{R}^{\rho, (2)}_{f} = \mathcal{F}^{\rho, (2)}_{f}(\epsilon) - \frac{1}{2}\Big(\mathcal{F}^{\rho, (1)}_{f}(\epsilon)\Big)^2 - 
  f^{\rho,(2)}(\epsilon)\mathcal{F}^{\rho, (1)}_{f}(2 
  \epsilon) - C^{\rho,(2)} + O(\epsilon).
\end{equation}
The form of the function $f^{\rho,(2)}\,(\epsilon)$ is inspired by the one that appears in BDS ansatz.  
Expanding $f^{\rho,(2)}(\epsilon)$ (see eq.~\ref{FR2} ~\cite{Brandhuber:2012vm}) as
\begin{eqnarray}
f^{\rho,(2)}(\epsilon) = \sum_{i=0}^2 \epsilon^i f^{\rho, (2)}_i
\end{eqnarray}
and demanding the finiteness of $\mathcal{R}^{\rho, (2)}_{f}$, we can uniquely fix $f^{\rho,(2)}_0$ and
$f^{\rho,(2)}_1$ but $f^{\rho,(2)}_2$ and $C^{\rho,(2)}$ can not be determined independently.    
Choosing 
\begin{equation}
C^{\rho,(2)} = 4 \zeta_4, 
\end{equation}
same for $\rho = BPS,{\cal K}$,
we obtain 
\begin{equation}
\label{FR3}
f^{\rho,(2)}(\epsilon) = -2 \zeta_2  + \epsilon \zeta_3 - \frac{1}{2} \epsilon^2 \zeta_4.
\end{equation}
Note that the $f^{\rho,(2)}$ as well as $C^{\rho,(2)}$ are independent of the operators as well as the
choice of external states.
This is a consequence of the universality of IR divergences in ${\cal N}=4$ SYM. 
The two loop finite remainder function, $\mathcal{R}^{\rho, (2)}_{f}$, can be expressed in terms of finite parts $\mathcal{F}^{\rho,(i),fin}_{f}$ in eq. (\ref{Reg11}) as

\begin{align}
\mathcal{R}^{\rho, (2)}_{f} =\,& \mathcal{F}^{\rho,(2),fin}_{f} - \left( \frac{1}{2}\mathcal{F}^{\rho,(1),fin}_{f} + 2K\right)\mathcal{F}^{\rho,(1),fin}_{f} + \frac{3}{4}A_{1}\left(f^{\rho,(2)}_{2} - \frac{3}{4}K\zeta_2\right) 
\nonumber\\
&- C^{\rho,(2)}-\frac{4}{3}\mathbf{H}^{\rho,(2)}_{f}\log\left( xyz \right),
\end{align}

where we have set $\mu^2 = -Q^2$. We have computed finite remainders for both the half-BPS and the Konishi operators for both the choices of 
external on shell states $f$, namely
$f=g \phi\phi$ and $f=\phi \lambda \lambda$.  The results, after setting $\mu^2=-Q^2 $, are listed below.
The  two loop finite remainder for the half-BPS operator turns out to be same for both choices of $f$: 
\begin{align} 
\label{FR4}
 \mathcal{R}^{\rm BPS, (2)}_{f} = \,&
 4 \Big( 2H(3,3,3,2; y) - H(2,3,2,3 ;y) + H(2,2,3,2 ;y) + H(2,2,1,0 ;y) 
\nonumber \\[-6mm] \nn &
 + 2 H(1,1,1,0 ;y) -    H(0,1,0,1 ;y) +   H(0,0,3,2 ;y) +   H(0,0,1,0 ;y)\Big) 
\nonumber \\[-6mm] \nn &
+ 8 \Big( H(2 ;y) +  H(3 ;y)\Big) \Big(H(0,0,1 ;z)- H(1,1,0 ;z)\Big) + 4\Big(H(2,3,2 ;y)  H(3 ;y) 
\nonumber \\[-6mm] \nn &
-  H(2,1,0 ;y)  H(2 ;y)  +   H(0,1,0 ;y)  H(1 ;y) - H(0,3,2 ;y)  H(0 ;y) 
\nonumber \\[-6mm] \nn &
-    H(1,0 ;y)  H(3,2 ;y) \Big) + 4  H(0,1 ;z)  \Big(H(2 ; y) + H(3;y)\Big)^2 
+  2H(1;z) \Big[H(3;y)
\nonumber \\[-6mm] \nn &
\times \Big(-H(0;y)^2+ 2  H(2;y)  H(3;y)  +  \frac{2}{3}  H(3;y)^2
 - 2  H(1,0;y) - 2 H(3,2;y)\Big) 
 \nonumber \\[-6mm] \nn &
 - 2 H(2,1,0;y)+ 2 H(2,2,3;y) + 2 H(3,0,0;y)\Big] -2 H(0;z)
 \Big[ H(0;y) \Big(H(2;y)^2 
 \nonumber \\[-6mm] \nn &
 -2 H(0,2;y)\Big)-2 H(2;y) H(1,0;y) +2  H(0;y) H(3,2;y)
 + 2  H(0,0,2;y) 
 \nonumber \\[-6mm] \nn &
 - 2  H(0,2,2;y)  -2 H(0,3,2;y) +2H(2,1,0;y) + H(1;z)\Big(H(2;y)^2 
\nonumber \\[-6mm] \nn &
+2H(2,0;y) +2H(3,0;y) +2 H(3,2;y)\Big)\Big] + 4 \zeta_{2} \Big[ 2 H(0,1;y) - 2 H(3,2;y)
\nonumber \\[-6mm] \nn &
 + H(1;y)^2 - 2 H(1;z) H(2;y) - H(2;y)^2 - 2 H(1;z) H(3;y) \Big].
\end{align}
The finite remainders for the Konishi operator for 
$f = g\phi\phi$ is given below
\begin{align}
\mathcal{R}^{{\cal K}, (2)}_{g\phi\phi}=  \, & \mathcal{R}^{\rm BPS, (2)}_{f}  - \zeta_{3}\frac{6y(-1 + y + z)}{(1 - z)^2}  + \zeta_{2}\Big [\frac{6y(-1 + y + z)}{(1 - z)^2}\Big(H(0; y) + H(0; z) \Big) 
\nonumber \\[-6mm] \nn &
 - \frac{3 H(1; y)}{z^2} (-1 + y)(-1 + y + 2z)  + \frac{3yH(1; z)}{z^2(1 - z)^2}\Big(10z^2(-1 + z) + y(-1 + 2z   
 \nonumber \\[-6mm]  \nn  &
 + 9z^2)\Big) + \frac{3 H(2; y)}{z^2(1 - z)^2} \Big((-1 + z)^4 + y^2(1 - 2z + 3z^2) + y(-2 + 6z - 8z^2 + 4z^3)\Big) \Big] 
 \nonumber \\[-6mm]  \nn  &
 - \frac{3 H(0; z)^2} {(1 - z)} \Big(y H(0; y) + H(2; y)(-1 + y + z)\Big) + H(0, 1; z)\Big[ H(3; y) \Big( \frac{3y^2} {z^2}  - 3 \Big)
 \nonumber \\[-6mm]  \nn  &
 - \frac{6y H(0; y)}{(1 - z)^2}(-1 + y + z) - \frac{ 3 H(2; y)}  {z^2(1 - z)^2} \Big((-1 + z)^4 + y^2(1 - 2z + 3z^2) + y(-2 
 \nonumber \\[-6mm]  \nn  &
 + 6z - 8z^2 + 4z^3)\Big)   \Big] + \frac{ 3 y H(0; z) H(0, 2; y)}{z^2(1 - z)^2} \Big(4(-1 + z)z^2 + y(-1 + 2z + 3z^2) \Big) 
 \nonumber \\[-6mm]  \nn  &
 +\frac{ 3H(1, 0; y) } {z^2} \Big[ -\frac{ H(0; z)} {(1 - z)^2} \Big((-1 + z)^2(-1 + 2z) + y^2(-1 + 2z + 3z^2) +  2y(1
\nonumber \\[-6mm]  \nn  &
-3z + z^2 + z^3)\Big)  +  H(1; z) (-1 + y)(-1 + y + 2z) + H(2;y)(-1 + y)(-1 + y 
\nonumber \\[-6mm]  \nn  &
+ 2z) \Big] - \frac{ 3 H(0; z) H(2, 0; y)} {z^2(1 - z)^2} \Big((-1 + z)^4 + y^2(1 - 2z - 3z^2) - 2y(1 - 3z + z^2 
\nonumber \\[-6mm]  \nn  &
+ z^3)\Big)+ H(3,2;y) \Big [H(0;y)\Big(3 - \frac{3y^2}{z^2}\Big) - \frac{3 H(0; z)} {z^2(1 - z)^2} \Big(4y(-1 + z)z^2 + z^2  
\nonumber \\[-6mm]  \nn  &
\times (-1 + z)^2 + y^2(-1 + 2z + 3z^2) \Big)\Big] + H(1 ; z) \Big[ \frac{3 y H(0; y)H(0; z)}{z^2(1 - z)^2} \Big(4(-1 + z)z^2  
\nonumber \\[-6mm]  \nn  &
+ y(-1+ 2z + 3z^2)\Big) - \frac{3 H(0; z)H(3;y)} { z^2(1 - z)^2 } \Big(4y(-1 + z)z^2 + (1 - z)^2z^2 + y^2(-1  
\nonumber \\[-6mm]  \nn  &
+ 2z + 3z^2) \Big)  - \frac{3H(2; y)H(3; y) } {z^2(1 - z)^2} \Big ((-1 + z)^4 + y^2(1 - 2z + 3z^2) + y(-2 + 6z  
\nonumber \\[-6mm]  \nn  &
- 8z^2 + 4z^3)\Big) + \frac{3}{2} H(3; y)^2 \Big(-1 + \frac{y^2 }{z^2} \Big) -  \frac{ 6yH(0, 3; y) }{(1 - z)^2 }  (-1 + y + z)+ H(3, 0; y)
\nonumber \\[-6mm]  \nn  &
\times \Big(3 - \frac{3y^2} {z^2} \Big) + \frac{ 3 H(3, 2; y) } {z^2(1 - z)^2} \Big((-1 + z)^4 + y^2(1 - 2z + 3z^2) +  y(-2 + 6z 
\nonumber \\[-6mm]  \nn  &
- 8z^2 + 4z^3)\Big)  \Big] -  \frac{6 y H(0, 0, 1; z)}{z^2(1 - z)^2} \Big (2  z^2 (-1 + z) +  y(-1 + 2z + 2z^2) \Big)
\nonumber \\[-6mm]  \nn  &
+ \frac{3 y H(0, 1, 0; y)}{z^2(1 - z)^2} \Big (2  z^2 (-1 + z) +  y(1 - 2z + 3z^2) \Big)
+ \frac{ 3y H(0, 1, 0; z)} {z^2(1 - z)^2} \Big(4(-1 + z)z^2 
\nonumber \\[-6mm]  \nn  &
+ y(1 - 2z + 3z^2)\Big) - \frac{ 3 H(0, 3, 2;y)}{z^2(1 - z)^2} \Big(2yz^2(-1 + z) + z^2(-1 + z)^2 + y^2(-1 + 2z 
\nonumber \\[-6mm]  \nn  &
+ z^2) \Big) - \frac{ 6y H(1, 0, 0; z) } {(1 - z)} - \frac {3y H(1, 0, 1; z)}{ z^2(1 - z)^2 } \Big(2z^2(-1 + z) + y(1 - 2z + 3z^2)\Big)
\nonumber \\[-6mm]  \nn  & 
- \frac{ 3 H(1, 1, 0; y)} {z^2} (-1 + y)(-1 + y + 2z)  + \frac{ 6y H(1, 1, 0; z)}{z^2(1 - z)^2} \Big(4(-1 + z)z^2 
\nonumber \\[-6mm]  \nn  &
+ y(-1 + 2z + 3z^2)\Big) + \frac{3 H(2, 1, 0;y)}{z^2(1 - z)^2} \Big((-1 + z)^2(-1 + 2z) + y^2(-1 + 2z + z^2) 
\nonumber \\[-6mm]  \nn  &
+  y(2 - 6z + 4z^2)\Big) - \frac{ 3H(2, 3, 2; y)}{z^2(1 - z)^2} \Big( (-1 + z)^4 + y^2(1 - 2z + 3z^2) + y(-2 
\nonumber \\[-6mm]  \nn  &
+ 6z - 8z^2 + 4z^3)\Big) + H(3, 3, 2;y) \Big(-3 + \frac{3y^2} {z^2} \Big) +\frac{1}{(1 - z)^4}  \Big[\zeta_2 \frac{ (-1 + z)^2}{z}
\nonumber \\[-6mm]  \nn  &
\times \Big(-52z(-1 + z)^2 + 6y^2(-1 + 5z) + y(9 - 48z + 39z^2)\Big) 
- 3y H(0; z)^2\Big(6y^3 
\nonumber \\[-6mm]  \nn  &
+ 12y^2(-1 + z) + 5y(-1 + z)^2 - (-1 + z)^3 \Big) + \frac{3(-1 + z)^3(-1 + 11z)}{z} H(0; z)
\nonumber \\[-6mm]  \nn  &
\times \Big(yH(0; y) + (-1 + y + z)H(2, y) \Big) -\frac{3(-1 + z)^2H(1, z)}{z^2} \Big[H(0; y)  y(-1 + y + z)
\nonumber \\[-6mm]  \nn  &
\times (-1 + z^2) + H(0; z) yz\Big(1 + 16z - 17z^2 - 2y(1 + 3z)\Big) + (-1 + z)H(3; y) 
\nonumber \\[-6mm]  \nn  &
\times \Big(-(y^2(1 + z)) + y(1 - z + 2z^2) + z(1 - 12z + 11z^2)\Big) \Big]  
\nonumber \\[-6mm]  \nn  &
- \frac{3yH(0, 1; z)} {z^2} (-1 + z)^2\Big(-1 + y + 2yz - 7z^2 + 5yz^2 + 8z^3 \Big) 
\nonumber \\[-6mm]  \nn  &
- \frac{3y H(0, 2;y)}{z^2} (-1 + z)^3(1 + z)(-1 + y + z)
\nonumber \\[-6mm]  \nn  &
-\frac{ 3 H(1, 0;y)}{z^2} (-1 + z)^3 \Big(8z^2(-1 + z) - y^2(1 + z) + y(1 + z - 4z^2)\Big) 
\nonumber \\[-6mm]  \nn  &
- \frac{3yH(2, 0;y)}{z^2}(-1 + z)^3(1 + z)(-1 + y + z) + \frac{3H(3, 2; y)}{z^2}(-1 + z)^3 \Big(y^2(1 + z) 
\nonumber \\[-6mm]  \nn  &    
+ z(-1 + 12z - 11z^2) + y(-1 + z - 2z^2)\Big)\Big] - \frac{12 y H(0;z)}{z(1 - z)^3} \Big(3y^3 + 6y^2(-1 + z) 
\nonumber \\[-6mm]  \nn  &
+ 5z(-1 + z)^2 + y(3 - 11z + 8z^2)\Big) - \frac{2}{z^2(1 - z)^2} \Big ( 9y^4 + 18y^3(-1 + z) 
\nonumber \\[-6mm]  \nn  &
+ 30yz(-1 + z)^2 - 53(-1 + z)^2z^2 + y^2(9 - 48z + 39z^2) \Big)
.
\end{align}
The finite remainders of the Konishi operator
for $f = \phi\lambda\lambda$ reads as
\begin{align}
\mathcal{R}^{{\cal K},(2)}_{\phi\lambda\lambda}  =  \,&\mathcal{R}^{\rm BPS, (2)}_{f} +  3\Big(  -9H(0,0,0,1;z) -2H(0,0,1,0;y) -5H(0,0,1,1;z) 
\nonumber \\[-6mm]  \nn  &
-3H(0,0,3,2;y)+3H(0,1,0,1;y) - 4H(0,1,0,1;z) + 2H(0,1,1,0;y) 
\nonumber \\[-6mm]  \nn  &
+ 2H(0,1,1,0;z)+H(0,2,1,0;y) - H(0,3,2,2;y) + H(0,3,3,2;y)  
\nonumber \\[-6mm]  \nn  &
- H(1,0,0,1;z)+ 2H(1,0,1,0;z) -2H(1,1,0,0;y) - 2H(1,1,0,1;z) 
\nonumber \\[-6mm]  \nn  &
- 4H(1,1,1,0;y)+H(2,0,1,0;y) - H(2,1,0,0;y) +H(2,1,1,0;y) 
\nonumber \\[-6mm]  \nn  &
- 3H(2,2,1,0;y)-2H(2,2,3,2;y)-H(2,3,0,2;y)-H(2,3,2,0;y) 
\nonumber \\[-6mm]  \nn &  
+ 3H(2,3,2,3;y)+2H(2,3,3,2;y) - 2H(3,3,2,2;y)-4H(3,3,3,2;y)\Big) 
\nonumber \\[-6mm]  \nn &  
+ 3\zeta_2\Big[ \frac{3}{2}\Big( H(1;z)+H(2;y) \Big)^2  - 2H(1;y)^2 + 4H(1;z)H(3;y) -4H(0,1;y) 
\nonumber \\[-6mm]  \nn &  
+3H(0,1;z) + H(2,1;y) + 4H(3,2;y)\Big] + 3H(0,0,1;z)\Big( 2H(0;y)+2H(0;z)
\nonumber \\[-6mm]  \nn &  
-5H(2;y)-4H(3;y) \Big) + 3H(0,1,1;z)\Big(  5H(0;y)-4H(3;y) \Big)+3H(1,0,1;z)
\nonumber \\[-6mm]  \nn &  
\times \Big( 2H(0;y)-2H(2;y)-3H(3;y) \Big) + 6H(1,1,0;z)\Big( H(3;y)-H(0;y) \Big)
\nonumber \\[-6mm]  \nn &  
+3H(0,1,0;y)\Big(  H(1;z)-3H(1;y)+H(2;y) \Big) + 6H(1,0,0;y)\Big(   H(0;z)
\nonumber \\[-6mm]  \nn &  
+2H(1;z)+2H(2;y) \Big) + 3H(0,0,2;y)\Big(  2H(0;z)+5H(1;z) \Big)-3H(2,0,0;y)
\nonumber \\[-6mm]  \nn &  
\times \Big( 2H(0;z)+5H(1;z) \Big) + 6H(2,1,0;y)\Big(  H(0;z)+H(1;z)+H(2;y) \Big)
\nonumber \\[-6mm]  \nn &  
+3H(2,3,2;y)\Big(  2H(0;y)-3 H(3;y)\Big) + 6H(3,2,2;y)\Big(  2H(0;y)-H(0;z) \Big)
\nonumber \\[-6mm]  \nn &  
+\frac{3}{4}H(1;z)^2\Big( 8H(3,0;y) + 6H(0,3;y)- 2H(3;y)^2-5H(0;y)^2  \Big)
\nonumber \\[-6mm]  \nn &  
+3H(1;z)\Big(   H(0,3,3;y) - 2H(2,2,3;y) - H(2,3,0;y) +2H(2,3,3;y) 
\nonumber \\[-6mm]  \nn &  
 -H(3,0,0;y) - 2H(3,3,2;y)   \Big) + 3H(0;y) \Big[  H(0;z)\Big(  H(2;y)^2-2H(0,2;y) 
\nonumber \\[-6mm]  \nn &  
+2H(3,2;y)  \Big) + H(1;z)\Big(  -5H(0,2;y)+2H(2,3;y)+4H(3,2;y) \Big) 
\nonumber \\[-6mm]  \nn &  
+2H(0,3,2;y) \Big] + 3H(1;z)H(0,3,0;y) - 3H(0,3,2;y)\Big( 2H(0;z)+H(1;z) \Big)
\nonumber \\[-6mm]  \nn &  
+3H(0,1;z)\Big[ \frac{1}{2}H(0;y)^2 -H(2;y)^2-\Big(4H(2;y) + 2H(3;y)\Big)H(3;y) 
\nonumber \\[-6mm]  \nn &  
+4H(0;y)H(2;y)+H(0,3;y)-H(1,0;y)-H(3,2;y) \Big] + 3H(1,0;y)
\nonumber \\[-6mm]  \nn &  
\times \Big(  H(3,2;y)+H(1;z)H(3;y)-2H(0;z)H(2;y) \Big)-15H(0,0;y)H(2,2;y) 
\nonumber \\[-6mm]  \nn &  
+ \frac{3}{2}H(1;z)\Big[  H(0;y)^2\Big(  H(3;y)-2H(0;z) \Big) -H(3;y)^2 \Big(  6H(2;y)+\frac{4}{3}H(3;y) \Big) 
\nonumber \\[-6mm]  \nn &  
+4H(0;z)H(3,0;y)+6H(3;y)H(3,2;y)\Big]  
\nonumber \\[-6mm]  \nn &  
+\frac{6z}{1-y}\zeta_3 - \frac{3\zeta_2}{1-y}\Big[  2z\Big(H(0; y) + H(0; z)\Big) -\frac{H(1;y)}{y(1-y-z)} \Big(  1 - 3z + 2z^2 
\nonumber \\[-6mm]  \nn &  
+ y^2(1 + 7z) + y(-2 - 4z + 6z^2) \Big) + \frac{H(1;z)}{(y+z)(1-y-z)}\Big( y^2(1 + 3z) 
\nonumber \\[-6mm]  \nn &  
+ z(5 - 3z - 2z^2) + y(-1 - 8z + z^2)  \Big) - \frac{H(2;y)}{y(y+z)}\Big((-1 + z)z + y^2(1 + 3z) 
\nonumber \\[-6mm]  \nn &  
- y(1 + 4z + 3z^2)\Big)\Big]  + \frac{3(y-5z)}{y+z}H(0;y)H(2;y)^2 + \frac{3(y-5z)}{y+z}H(1;z)^2\Big( H(0;y)
\nonumber \\[-6mm]  \nn &  
-H(3;y)\Big) + \frac{3H(0,1;z)}{y+z}\Big[ -\frac{2H(0;y)}{1-y}\Big( 3y^2 + z(-2 + 3z) + y(-3 + 5z) \Big)  
\nonumber \\[-6mm]  \nn &  
+ \frac{H(2;y)}{y(1-y)}\Big(  6y^3 - (-1 + z)z - y^2(7 + 3z) + y(1 + 6z + 5z^2)\Big) 
\nonumber \\[-6mm]  \nn &  
-\frac{H(3;y)}{y(1-y-z)}\Big( z(-1 + z)(8y+ z) + y^2(1 + 7z) \Big)\Big] + H(0,2;y)\Big[  \frac{6H(0;y)}{1-y} (-1
\nonumber \\[-6mm]  \nn &  
+y+6z) -\frac{3H(0;z)}{(1-y)(y+z)(1-y-z)}\Big( 4y^3 + y^2(-7 + 9z) + z(7 - 5z - 2z^2) 
\nonumber \\[-6mm]  \nn &  
+ y(3 - 16z + 3z^2)  \Big) + \frac{6(y-5z)}{y+z}H(1;z) \Big] - \frac{3H(1,0;y)}{y(y+z)(1-y-z)}\Big[ H(0;z)\Big(4y^3
\nonumber \\[-6mm]  \nn &  
 - (-1 + z)z + y^2(-5 + 9z) + y(1 - 6z + 5z^2)  \Big)  + \Big(H(1; z) + H(2; y)\Big)\Big( 4y^3
\nonumber \\[-6mm]  \nn &  
 - (-1 + z)z + y^2(-5 + 11z) + y(1 - 8z + 7z^2) \Big) \Big] + \frac{3H(2,0;y)}{y+z}\Big[ \frac{H(0;z)}{y(1-y)}\Big( 4y^3
\nonumber \\[-6mm]  \nn &  
 - (-1 + z)z + y^2(-5 + 3z) + y(1 - 8z - 3z^2)  \Big) + 2(y-5z)H(1;z) \Big]
\nonumber \\[-6mm]  \nn &  
-\frac{3H(3,2;y)}{y(1-y)(1-y-z)}\Big[  H(0;y)\Big( 4y^3 + y^2(-7 + z) + (-1 + z)z - 3y(-1 + z^2)\Big)  
\nonumber \\[-6mm]  \nn &  
+H(0;z)(1-y)\Big(  4y^2 - (1-z)(3y-z)  \Big) \Big] + H(1;z)\Big[ \frac{3H(0;y)^2}{1-y}(-1+y+6z)
\nonumber \\[-6mm]  \nn &  
+\frac{3H(0;y)H(0;z)}{(y+z)(1-y-z)}\Big( 4y^2 + 5(-1 + z)z + y(-3 + 9z)  \Big) 
\nonumber \\[-6mm]  \nn &  
- \frac{6z(1+z)}{(1-y)(y+z)}H(0;z)H(2;y) - \frac{3H(0;z)H(3;y)}{y(1-y-z)}\Big( 4y^2-(1-z)(3y-z) \Big) 
\nonumber \\[-6mm]  \nn &  
 + \frac{3H(2;y)H(3;y)}{y(1-y)(y+z)}\Big( 6y^3 - (-1 + z)z - y^2(7 + 3z) + y(1 + 4z + 3z^2)  \Big)
\nonumber \\[-6mm]  \nn &  
-\frac{3H(3;y)^2}{2y(y+z)(1-y-z)}\Big( 8y(-1 + z)z + (-1 + z)z^2 + y^2(1 + 7z)  \Big)
\nonumber \\[-6mm]  \nn &  
-\frac{6H(0,3;y)}{1-y} (-3+3y+2z)  - \frac{3H(3,0;y)}{y(1-y)(1-y-z)} \Big(  4y^3 + y^2(-7 + z) 
\nonumber \\[-6mm]  \nn &  
+ (-1 + z)z - 3y(-1 + z^2)  \Big) -  \frac{3H(3,2;y)}{y(1-y)}\Big(  1 + 4y^2 - z + y(-5 + 3z) \Big)  \Big]
\nonumber \\[-6mm]  \nn &  
-\frac{6H(0,0,1;z)}{(1-y)(y+z)(1-y-z)}\Big( -3y^2 + y^3 + 2y(1 + z) + z(-2 + z + z^2)  \Big) 
\nonumber \\[-6mm]  \nn &  
 -\frac{6H(0,0,2;y)}{1-y}(-1+y+6z)+\frac{3}{(1-y)(y+z)(1-y-z)}\Big[   H(0,1,0;y)\Big( 6y^3 
\nonumber \\[-6mm]  \nn &  
+ y^2(-11 + 21z) + 3z(1 - 5z + 4z^2) + y(5 - 24z + 27z^2) \Big) + H(0,1,0;z)\Big( 4y^3 
\nonumber \\[-6mm]  \nn &  
+ 7y^2(-1 + z) + z(1 - 3z + 2z^2) + y(3 - 8z + 5z^2)  \Big)\Big] -\frac{6(y-5z)}{y+z}H(0,1,1;z)
\nonumber \\[-6mm]  \nn &  
+\frac{3H(0,3,2;y)}{y(1-y)(1-y-z)}\Big(10y^3 + (-1 + z)z + y^2(-19 + 11z) + y(9 - 10z + z^2)  \Big)
\nonumber \\[-6mm]  \nn &  
- \frac{6(-1+y+6z)}{1-y}H(1,0,0;y) - \frac{3H(1,0,1;z)}{(1-y)(y+z)(1-y-z)}\Big( 6y^3 + y^2(-11 + 5z) 
\nonumber \\[-6mm]  \nn &  
+ y(5 - 2z + z^2) + z(-3 + z + 2z^2) \Big) + \frac{3H(1,1,0;y)}{y(1-y)(1-y-z)}\Big( 1 - 3z + 2z^2 
\nonumber \\[-6mm]  \nn &  
+ y^2(1 + 7z) + y(-2 - 4z + 6z^2)  \Big) - \frac{6H(1,1,0;z)}{(1-y)(y+z)(1-y-z)}\Big( y^2(1 + z) 
\nonumber \\[-6mm]  \nn &  
- y(1 + 3z) - z(-2 + z + z^2)   \Big)  + \frac{6(-1+y+6z)}{1-y}H(2,0,0;y)
\nonumber \\[-6mm]  \nn &  
+\frac{3H(2,1,0;y)}{y(y+z)(1-y-z)}\Big( 10y^3 - (-1 + z)z + y^2(-11 + 19z) + y(1 - 10z + 9z^2) \Big)
\nonumber \\[-6mm]  \nn &  
+\frac{3H(2,3,2;y)}{y(1-y)(y+z)}\Big( 6y^3 - (-1 + z)z - y^2(7 + 3z) + y(1 + 4z + 3z^2)  \Big)
\nonumber \\[-6mm]  \nn &  
-\frac{6(y-5z)}{y+z}H(3,2,2;y) - \frac{3H(3,3,2;y)}{y(y+z)(1-y-z)}\Big( 8y(-1 + z)z + (-1 + z)z^2 
\nonumber \\[-6mm]  \nn &  
+ y^2(1 + 7z)  \Big) - \frac{\zeta_2}{1-y}\Big(52(1-y)+12z\Big) -\frac{18z^2}{(1-y)^2}H(0;y)^2
\nonumber \\[-6mm]  \nn &  
-\frac{18z^2}{(y+z)^2}\Big( H(1;z)+H(2;y) \Big)^2 + H(1;z)\Big[  \frac{6H(0;y)}{(1-y)(y+z)}\Big(  y^2 + y(-1+z) 
\nonumber \\[-6mm]  \nn &  
+ z(-1+6z) \Big) - 24H(0;z) + \frac{6(y-z)}{y+z}H(3;y) \Big] + \frac{6(5y+3z)}{y+z}H(0,1;z)   
\nonumber \\[-6mm]  \nn &  
+\frac{6H(0;y)H(2;y)}{(1-y)(y+z)}\Big(  y^2 + y(-1+z) + z(-1+6z) \Big)  - \frac{6(-1+y+2z)}{1-y}H(1,0;y)
\nonumber \\[-6mm]  \nn &  
 + \frac{6(y-z)}{y+z}H(3,2;y)-\frac{24z}{1-y}H(0;y) + \frac{24z}{y+z}\Big(  H(1;z) + H(2;y) \Big) + 106
.
\end{align}
Note that ${\cal R}^{BPS,(2)}_f$ for both choices of $f$
contain terms of uniform weight four resulting from HPLs and Riemann zeta function($\zeta_n$s) with
rational coefficients, i.e., no dependence on the scaling variables $y,z$. 
This is exactly the case for the finite terms of FFs after subtracting
the Catani's subtraction operators and this is not to do with the exponentiation.  
For the Konishi operator, the results contain HPLs of different weights.  Interestingly,
HPLs with highest weight in the Konishi when $f=g \phi \phi$ coincide exactly with
the result of BPS.  This is not the case for $f=\phi\lambda\lambda$.   
Like the BPS, the coefficients of leading HPLs in ${\cal R}^{{\cal K},(2)}_f$  
are rational terms with no $y,z$ dependence for both choices of $f$.  
The reason is still unclear. 
Using the fortran routines of Gehrmann and Remiddi \cite{Gehrmann:2001jv}
we have compared the finite remainder of the half-BPS operator presented in eq. (\ref{FR4})  against the compact result
presented in the (eq. (4.32) of \cite{Brandhuber:2012vm}) and found excellent agreement 
within the numerical accuracy.  
The numerical comparison for various values of $(x,y,z)$ is presented in the table above.
\begin{table}
\begin{center}
\label{LT6}
\begin{tabular}{ | c | c | c | }
\hline 
&& \\
($x,y,z$) & $\mathcal{R}^{(2)}_{3}$ \cite{Brandhuber:2012vm}  &  $\mathcal{R}^{{\rm BPS},(2)}_f$  \\[1ex]
&& \\
\hline \hline 
 (1/7,\, 1/7,\, 5/7)        & -0.08565426912931533 &  -0.08565426912932168  \\[1ex]
 \hline 
 (7/9,\, 1/9,\, 1/9)        &-0.06628498978762110  & -0.06628498978762294       \\[1ex]
 \hline
 (4/11,\, 1/11,\, 6/11)     &-0.09538372586521941  &-0.09538372586522141     \\[1ex]
\hline
(20/33,\, 7/33,\, 6/33)     &-0.11365299782099999 &-0.11365299782099962    \\[1ex]
\hline
(10/19,\, 2/19,\, 7/19)     &-0.10323613406746748 &-0.10323613406746739    \\[1ex]
\hline
(8/17,\, 4/17,\, 5/17)      &-0.13812300982989027 &-0.13812300982989134    \\[1ex]
\hline
(75/99,\, 1/99,\, 23/99)    &-0.01926355459500375  &-0.01926355459500363     \\[1ex]
\hline
(67/79,\, 5/79,\, 7/79)    &-0.04241579611131384 & -0.04241579611131386     \\[1ex]
\hline
 \end{tabular}
\end{center} 
\caption{Numerical comparison of eq. (4.32) of \cite{Brandhuber:2012vm} against our eq. (\ref{FR4})} 
\end{table}

\section{Universality of leading transcendental terms}
\label{sec:LT}
Based on the supersymmetric extensions of BFKL and DGLAP evoultion equations, 
Kotikov and Lipatov ~\cite{Kotikov:2000pm,Kotikov:2002ab,Kotikov:2004er} (see also \cite{Kotikov:2001sc,
Kotikov:2006ts}) conjectured maximum transcendentality principle which implies that the anomalous
dimensions of leading twist two operators in ${\cal N}=4$ SYM contain uniform 
transcendental terms which are related to those in the 
corresponding QCD results \cite{Moch:2004pa,Vogt:2004mw}.  
The transcendentality weight $n$ is defined by terms such as  $\zeta(n)$, $\epsilon^{-n}$ and
weight of the HPLs that appear in  
the perturbative calculations.   
In ${\cal N}=4$ SYM, scattering amplitudes of certain type \cite{Bern:2006ew,Naculich:2008ys}, FFs of the BPS type operators \cite{Bork:2010wf,Gehrmann:2011xn,Brandhuber:2012vm,Eden:2012rr}, 
light-like Wilson loops \cite{Drummond:2007cf,Drummond:2013nda} 
and correlation functions \cite{Eden:2012rr,Drummond:2013nda,Basso:2015eqa,Goncalves:2016vir} surprisingly 
exhibit this property.   
In \cite{Brandhuber:2012vm}, the two loop three point MHV FFs of the half-BPS operators were shown to
have uniform transcendental terms in the finite reminder functions.  Using the method of symbology  
which uses coproducts, the finite reminder function of the half-BPS operator at two loops 
was expressed in terms of transcendental weight four functions which agreed numerically with their
result obtained using the unitarity based approach. Similar results for non-protected operators can be found in ~\cite{Loebbert:2015ova,Brandhuber:2016fni,Loebbert:2016xkw}.

Unlike ${\cal N}=4$ SYM, QCD results contain terms of all transcendental weights in addition to
rational terms (zero transcendentality).  But surprisingly, leading transcendental terms of anomalous
dimensions of twist two operators, several FFs in QCD when $C_A=C_F=N$ and 
$T_f n_f=N/2$ coincide with the corresponding ones in ${\cal N}=4$ SYM.  
For example, the leading transcendental terms of 
the amplitude for Higgs boson decaying to three on-shell gluons in 
QCD ~\cite{Gehrmann:2011aa,Koukoutsakis:2003nba} are related to 
the two loop three point MHV FFs of the half-BPS operators ~\cite{Brandhuber:2012vm}.
Similarly, the same is true for two point FFs of quark current operator, scalar and pseudo scalar operators constructed
out of gluon field strengths, energy momentum tensor of the QCD up to three loops. 
In section \ref{sec:IR}, we determined both $K$ and $\textbf{H}^{(2)}_f$ appearing in Catani's IR subtraction operator by 
demanding the finiteness of ${\cal F}^{\rho,(n),fin}_f$, $n=1,2$. 
In QCD the single pole term in $\textbf{I}^{(2)}_f$ is denoted by $\textbf{H}^{(2)}_{i}$, where $i = q,\bar{q},g$. 
This was shown in ~\cite{Ravindran:2004mb,Ravindran:2005vv} to be related to the  the universal cusp ($A_{i}$) ~\cite{Vogt:2000ci, Vogt:2004mw}, 
 collinear ($B_{i}$)  and soft ($f_{i}$) ~\cite{Vogt:2004mw,Ravindran:2005vv} anomalous dimensions
  in the following way 
\begin{equation}
\label{QCDLT}
\textbf{H}^{(2)}_{i} = \frac{1}{\epsilon} \Big(\frac{1}{8} B_{i}^{(2)} + \frac{1}{16} f_{i}^{(2)} - \frac{1}{32C_{i}}  B_{i}^{(1)} A_{i}^{(2)} -  \frac{3}{16} C_{i} \beta_{0}\zeta_{2}\Big)\,,
\end{equation}
where $C_{q} = C_{F}$, $C_g = C_{A}$ and $\beta_0$ is coefficient of QCD $\beta-$function. Also the quantity $K$ that appears in $\textbf{I}^{(2)}_f$ is related to $A^{(2)}_i$. 
We find $\textbf{H}^{(2)}_{i}  = -\frac{1}{4\epsilon} N^2\zeta_3$ after setting the QCD colour factors to N and taking the leading transcendental ($\text{LT}$) terms of $A_{i}$, $B_{i}$, $f_{i}$ in eq. \eqref{QCDLT}.
In $\mathcal {N} = 4$ SYM for both choices of external states $f$, $\textbf{H}^{(2)}_{f} = \sum_{i=\{f\}}\frac{1}{N^2} \textbf{H}^{(2)}_{i} $(see eq. \eqref{eq:KH}).
Here, the factor $1/N^2$ that appears in front of $\textbf{H}^{(2)}_{i}$ is due to the definition of `t Hooft coupling constant given in eq. (\ref{eq:tHcoup}).    
The three point FFs of the half-BPS operator presented up to two loop level for $f=g\phi\phi$ and  $f=\phi\lambda\lambda$ contain
terms with uniform transcendentality of weight four.  On the other hand, the FFs of the Konishi operator for both
the choices of $f=g\phi\phi,\phi\lambda\lambda$ do not show this behaviour.  But interestingly,
the leading transcendental terms of it for only $f=g\phi\phi$ agree exactly with the result of the BPS operator.
For the choice $f=\phi\lambda\lambda$, this is not the case as the type of fields in the operator do not 
coincide with the external states.  In summary, up to two loops, we find
\begin{eqnarray}
  \label{LTN4}
  \mathcal {F}_{g\phi\phi}^{{\cal  K}, {(2)}} \Big |_{LT} =  \,\,\mathcal {F}_{ f }^{\rm BPS,{(2)}}  \,\, , \quad \quad
  \mathcal {F}_{\phi\lambda\lambda}^{{\cal  K}, {(2)}} \Big |_{LT} \neq  \,\, \mathcal {F}_{ f }^{\rm BPS,{(2)}} \,\, .
 \end{eqnarray}
The above relations hold for finite remainders of FFs of the BPS and the Konishi operator up to  
two loop level.  
\begin{eqnarray}
\label{LT7}
\mathcal{R}^{\rm BPS, (2)}_{g\phi\phi} =  \,\, \mathcal{R}^{\rm BPS, (2)}_{\phi\lambda\lambda} \,\,, \quad \quad
\mathcal{R}^{{ \cal K}, (2)}_{g\phi\phi}|_{LT}  = \,\,\mathcal{R}^{\rm BPS, (2)}_f\,\,,  \quad \quad
\mathcal{R}^{{ \cal K}, (2)}_{\phi\lambda\lambda}|_{LT}  \neq  \,\, \mathcal{R}^{\rm BPS, (2)}_f \,\, .
\end{eqnarray}
The reason for the equality between the leading transcendental terms of the Konishi and the BPS for $f=g \phi \phi$ 
could possibly be attributed to the fact that fields in the operator and those in
the external states coincide.  But one requires to do this exercise with more operators to confirm  
this reasoning.  

\section{Discussion and Conclusions}
\label{sec:conclusion}
In this paper we have computed the three point on-shell 
form factors for the half-BPS and the Konishi operators 
up to two  loop order in perturbative expansion by employing the Feynman diagrammatic methods. 
We have regularized all the divergences using the $\overline{DR}$ scheme.  
We have also repeated the computation by regularising in the FDH scheme and 
compared against the one computed in the $\overline{DR}$ after  properly modifying the FDH scheme 
for space-time  dependent operators.  We find  
complete agreement between the two approaches.
The  half-BPS operator is protected by supersymmetry and hence 
it does not get any quantum corrections to any order in the perturbative expansion. On the other hand, the scaling 
dimension  of the Konishi operator receives corrections at each order in perturbation theory. 
The ultraviolet divergence of the form factor with the Konishi operator insertion are removed 
by introducing an overall renormalization constant, these constants are function of the coupling constant
 and the anomalous dimension, where the
latter has been computed up to five loop order in perturbation theory. These matrix elements still 
suffer from divergences coming from the IR sector which are of universal nature as 
predicted  ~\cite{Catani:1998bh,Sterman:2002qn,Becher:2009cu,Gardi:2009qi} 
in the context of QCD.  The subtraction operators  which  encapsulate these predictions in $SU(N)$
gauge theory are given in terms
 of the colour factors $C_{A}, C_{F}, n_f T_f$.  In order to get the corresponding subtraction 
operators in ${\cal N} =4$ SYM (see eq. (\ref{Reg9})) we have put $\beta =0$.
 The two unknown quantities,  $K$ and $\textbf{H}^{(2)}_{f}(\epsilon)$, appearing in the
 subtraction operators have been computed by demanding that the subtraction operators 
 cancel all the IR divergences in the three point FFs. 
We find out that $K$ and $\textbf{H}^{(2)}_{f}(\epsilon)$  are the highest transcendental parts of the
corresponding expressions in QCD when proper
colour assignments are performed. In this context, we want to mention about some observations 
that we have made in our earlier work ~\cite{Ahmed:2016vgl}. The Sudakov FF in $\cal{N} = $ 4 SYM
was shown to satisfy the 
KG-equation, whose solution can be expressed  in terms of collinear, soft and cusp
anomalous dimensions. We have observed that these quantities are also the highest transcendental 
parts of the corresponding expressions in QCD, on assigning proper colour factors.
Also  the coefficient of the
single pole at each order in perturbative expansion coincides with  the highest transcendental part of $(2 B_{i}+ f_{i}) $. 
The UV and IR finite three point FFs, $ \mathcal{F}^{{\cal{K}},(2),\text{fin}}$ 
show some interesting features. For the half-BPS operator the FF results at one and two loop orders in 
perturbative expansion for both external states
 contain only highest transcendental terms (two and four respectively), expressed fully in terms of HPLs and $\zeta_2$
 (see  eqs. (\ref{Reg21a}, \ref{Reg11a})).
The coefficients of these HPLs and $\zeta$'s are not functions of  any kinematical invariants.
It is interesting to note that the Sudakov FF results for the above  operator was also found not to depend 
on the nature of the  external legs. In contrast to the half-BPS operator, the  three point FF result for the Konishi operator
contains both highest as well as lower transcendental terms. When computed between $g\phi\phi$ external  state
the FF result for the Konishi operator has  the same highest  transcendental contribution as that of  the half-BPS operator.
But for the same operator with $\phi \lambda \lambda$ external state, there is 
no resemblance of LT terms with the BPS counterpart.  

It is well known that the IR divergences of MHV amplitudes have an exponential 
structure (BDS ansatz). Such a behavior is reminiscent of the 
factorisation properties of gauge theory amplitudes. However we can also see this kind of exponentiation 
in FFs.
 By exponentiating the one loop IR singularities following BDS like exponentiation, we compute the 
finite remainder functions $ \mathcal{R}^{\rho, (2)}_{f}$ (see eq. \ref{FR2}).
We have presented the finite remainder functions  of the three point FFs for both the
operators up to second order in perturbative expansion for both choices of $f$. 
The authors in ~\cite{Brandhuber:2014ica} have computed three point function of the half-BPS
operator between the states $f=g \phi\phi$ using   
unitarity based techniques and presented their result in compact form using the method of
symbols that uses coproducts.  
Our result on the half-BPS for $f=g\phi\phi$ presented in terms of HPLs numerically agree with
theirs presented in terms of classical polylogarithms for the finite remainder function. 

The leading transcendental terms of the remainder 
function for the Konishi operator computed between $g\phi\phi$ external state coincide with 
the corresponding one of the half-BPS.
   However, for the $\phi \lambda \lambda$ 
external state we find that the expressions for the Konishi and the half-BPS operators do not have anything
common.   In this context, it is worth mentioning 
about the minimal FF results for a non-protected operator  presented in ~\cite{Loebbert:2015ova}. 
By computing the minimal FF in SU(2) sector, the authors demonstrate that the remainder functions do not have 
uniform transcendentality. In addition to this, they also report that the maximum transcendental terms
of the non protected operator are same as  the corresponding FF results for the BPS operator.
We find the situation for the Konishi is identical when computed between $g \phi\phi$ state, but not for
$\phi \lambda \lambda$.  The reason for the agreement for $g \phi\phi$ state could be 
due to the fact that both the Konishi operator and the external
states have two scalar fields in common.  To explore this further, we need more data on other unprotected operators.  

\section{Acknowledgements}
We would like to thank T.~Gehrmann, J.~Henn,  R.N.~Lee, P.~Nogueira  and A.~Tripathi for useful discussions and also
thank T.~Ahmed and N.~Rana for carefully going over the manuscript and providing useful suggestions.
\appendix
\section{Harmonic Polylogarithms}
\label{sec:appA}
We briefly describe the definition and properties of HPL and 2dHPL that we have used in
our paper.
The Harmonic Polylogarithm of weight w and  comprising of w-dimensional vector $\vec{m}_w$ of parameters
is represented by $H(\vec{m}_w ; y)$, where y is the argument of the function. The elements of  $\vec{m}_w$
$\in \{-1,0,1\}$ through which the following rational functions are expressed
\begin{equation}
\label{HP1}
f(-1;y) \equiv \frac{1}{1+y}, \qquad f(0;y) \equiv \frac{1}{y}, \qquad  f(1;y) \equiv \frac{1}{1-y}.
\end{equation}
These functions are related to $H(\vec{m}_w ; y)$ via
\begin{equation}
\label{HP2}
\frac{d}{dy} H(\vec{m}_w ; y) = f(\vec{m}_w;y).
\end{equation}
HPLs of weight 1 ($w=1$) follow from \eqref{HP1} and \eqref{HP2}
\begin{equation}
H(-1 ; y) = \log(1+y), \qquad  H(0; y) = \log(y), \qquad  H(1 ; y) = -\log(1-y).
\end{equation}
For $w>1$ $H(m, \vec{m}_w ; y)$ is given by
\begin{equation}
\label{eq:a3}
H(m, \vec{m}_w ; y) \equiv \int _{0}^{y} f(m;x) H(\vec{m}_w ; y),  \qquad  m  \in \{-1,0,1\}
\end{equation}
Moreover the higher dimensional HPLs are defined following the eq.~(\ref{eq:a3}) for the new elements
{2,3} in $\vec{m}_w$, representing a new class of rational functions
\begin{equation}
f(2;y) ~\equiv~ f(1-z;y) ~\equiv~ \frac{1}{1-y-z}, \qquad
f(3;y) ~\equiv~ f(z;y)   ~\equiv~ \frac{1}{y+z} ;
\end{equation}
correspondingly the weight-1 ($w = 1$) two-dimensional HPLs are given by
\begin{equation}
H(2,y)  ~\equiv~ -\text{ln}\Big(1-\frac{y}{1-z} \Big), \qquad
H(3,y)  ~\equiv~  \text{ln}\Big(\frac{y+z}{z}\Big).
\end{equation}
The Shuffle algebra that the HPLs follow is defined as follows:  
a product of two HPL with weights $w_1$ and $w_2$ of the same argument y is a combination of
HPLs with weight $(w_1 + w_2)$ and argument y, such that all possible permutations of the elements of  $\vec{m}_{w_{1}}$
and $\vec{m}_{w_{2}}$ are considered preserving the relative orders of the elements of  $\vec{m}_{w_{1}}$ and $\vec{m}_{w_{2}}$ ,
that is
\begin{equation} 
\label{B:6}
   H(\vec{m}_{w_1};y)~H(\vec{m}_{w_2};y) ~~ =
    \sum_{ \scriptscriptstyle{ \mathsmaller{ \vec{m}_{w} = \vec{m}_{w_1} \biguplus \vec{m}_{w_2} }  } } 
    H(\vec{m}_{w};y)
\end{equation} 
\section{Results for finite part of form factors }
For the half-BPS, and the Konishi operators at one loop, we find
\begin{eqnarray}
\label{Reg21a}
\mathcal{F}^{\rm BPS,(1),\text{fin}}_f & = & -\zeta_2-H(0;y)\Big (H(0;z)-H(1;z)\Big ) + H(2;y)\Big (H(0;z)+H(0;y)\Big )
\nonumber\\
 &&- H(1;z)\Big (2H(3;y) + H(0;z)\Big ) - 2H(1,0;y) - 2H(3,2;y) \, ,
\nonumber\\
\mathcal{F}^{\mathcal{K},(1),\text{fin}}_{g\phi\phi}  &= & \mathcal{F}^{\rm BPS,(1),\text{fin}}_f+ \frac{6y(1-y-z)}{(1-z)^2}H(0;z)
-\frac{2}{z(1-z)}\Big( 3y^2 - (3y-7z)(1-z)\Big) \, ,
\nn
\mathcal{F}^{\mathcal{K},(1),\text{fin}}_{\phi \lambda\lambda}  &= &\mathcal{F}^{\rm BPS,(1),\text{fin}}_f -3H(0;y)H(1;z) + 3H(0,1;z) -3H(0;y)H(2;y) 
\nn &&
+3H(1,0;y) +3H(1;z)H(3;y)+ 3H(3,2;y)+\frac{6z}{1-y}H(0;y)
\nn &&
-\frac{6z}{y+z}\Big (H(1;z) + H(2;y)\Big ) - 14.
\end{eqnarray}

The two loop finite result  for FF for the half-BPS operator is presented below 
\begin{align}
\label{Reg11a}
\mathcal{F}^{{\rm BPS}, (2),\text{fin}}_{f} = \, &  \frac{49}{20} \zeta_2^2 + 
\zeta_3 \Big ( H(1; z) + H(2; y) -H(0; y) - H(0; z) \Big )
+ \zeta_2 \Big [ 
  3 H(0; z) 
  \nonumber \\[-6mm] \nn &
\times  \Big (  H(1; z) +  H(0;y)  -  H(2;y)  \Big ) 
 + 4 \Big (H(1;y)^2 -  H(2;y)^2\Big)  
 - H(1; z)  
 \nonumber \\[-6mm] \nn &
\times\Big( 3 H(0; y) + 8 H(2; y) + 2H(3; y) \Big )    
 - 2 H(3, 2 ;y) - 3 H(0;y) H(2;y) 
 \nonumber \\[-6mm] \nn &
 + 6 H(1, 0; y)+  8 H(0, 1; y) \Big ] + H(0;z)\Big [ -H(0; y)^2H(1; z) 
- H(0;y)  
\nonumber \\[-6mm] \nn &
\times\Big (H(2; y)^2- 2H(0, 2; y)  + 2H(3, 2;y) \Big ) - 
    2\Big (H(1; z)\Big (H(2; y)^2        
- H(0, 3; y) 
\nonumber \\[-6mm] \nn &
+ H(2 ;y)H(3; y) - H(1, 0;y) + 2H(2, 0; y) + H(3, 0; y) + H(3, 2; y)\Big )  
\nonumber \\[-6mm] \nn & 
-H(2; y) H(1, 0;y) + H(0, 0, 2; y)  - H(0, 1, 0;y) - 2H(0, 2, 2;y) 
\nonumber \\[-6mm] \nn &
- 2H(0, 3, 2;y) 
   -   2H(1, 0, 0;y) + H(2, 0, 0;y) +  2H(2, 1, 0;y) + H(2, 3, 2;y) 
  \nonumber \\[-6mm] \nn &
   + 2H(3, 2, 2;y) \Big )\Big ]    
  +  \frac{1}{2} H(0,z)^2
     \Big (H(0; y)^2 + H(2;y)^2 - 2 H(0;y) H(2;y)\Big ) 
     \nonumber \\[-6mm] \nn &
     + H(1;z) \Big [    
     -2H(0;y)^2 H(3; y) + 8H(2; y)H(3; y)^2 +\frac{4}{3} H(3;y)^3
     \nonumber \\[-6mm] \nn &
     +2H(0; y) \Big (H(0, 2; y)  - 2H(3, 2; y)\Big ) - 8H(3; y)H(3, 2;y) -
    2H(0, 0, 2; y) 
    \nonumber \\[-6mm] \nn &
    - 2H(0, 1, 0; y)  - 2H(0, 2, 3; y) - 4H(1, 0, 0; y) +
    2H(2, 0, 0;y) 
    \nonumber \\[-6mm] \nn &
    - 2H(2, 0, 3; y) - 4H(2, 1, 0; y)    
+ 4H(2, 2, 3; y)  -2H(2, 3, 0; y) 
\nonumber \\[-6mm] \nn &
- 8H(2, 3, 3; y) + 4H(3, 0, 0; y) + 8H(3, 3, 2; y)\Big ]  
+ \frac{1}{2}H(1;z)^2 \Big ( H(0; y)^2   
\nonumber \\[-6mm] \nn &
+ 4H(3; y)^2 - 4H(0;y) H(3;y)\Big ) + 2 H(0;y)\Big (
H(0, 0, 1; z) + H(0, 1, 0;z) 
\nonumber \\[-6mm] \nn &
- H(0, 1, 1;z) - 2 H(0, 3, 2;y) +  H(1, 0, 0; z) 
- H(1, 0, 1; z) - H(1, 1, 0; z) 
\nonumber \\[-6mm] \nn &
- H(2,3,2;y) - 2H(3,2,2;y) \Big ) + 8 H(1;y) H(0, 1, 0; y) + H(2;y) 
\nonumber \\[-6mm] \nn &
\times \Big [ 8H(3; y)H(0, 1;z) 
+ 6H(0, 0, 1;z) - 2 \Big (H(0, 1, 0; z) 
+ H(1, 0, 0; z)  
\nonumber \\[-6mm] \nn &
+ 4H(1, 1, 0; z) + 2H(2, 1, 0; y) + H(0,1,0;y) + 2 H(1,0,0;y)\Big )\Big ]
\nonumber \\[-6mm] \nn &
+ 4  H(2;y)^2  H(0, 1; z) 
+ 4 H(3;y) \Big ( 2H(0, 0, 1;z) + H(0, 1, 1;z)  - H(1, 1, 0; z) 
 \nonumber \\[-6mm] \nn &
 + H(1, 0, 1; z) + 2H(2, 3, 2; y) \Big ) + 4 H(3;y)^2 H(0, 1; z)  
 \nonumber \\[-6mm] \nn &
 + 2 \Big (H(0, 0;z)H(1, 1; z) + H(0, 0; y)H(2, 2; y) + 2H(0, 0, 1, 0;y) 
 \nonumber \\[-6mm] \nn &
 + 2H(0, 0, 3, 2;y) - 4H(0, 1, 0, 1;y) - 4H(0, 1, 1, 0;y) +  4H(1, 1, 0, 0; y) 
 \nonumber \\[-6mm] \nn &
 + 4H(1, 1, 1, 0; y) + 2H(2, 2, 1, 0;y)+ 2H(2, 2, 3, 2; y) 
 \nonumber \\[-6mm] \nn &
 - 4H(2, 3, 2, 3; y) - 4H(2, 3, 3, 2; y) + 4H(3, 3, 2, 2; y) + 4H(3, 3, 3, 2; y) \Big ) 
.
\end{align}

The expression above contains only HPLs (see  appendix \ref{sec:appA} ) and zeta functions. We have used several 
identities among HPLs to simplify the expressions.
The FF expression for the Konishi operator
sandwiched between $g\phi\phi$ states reads as
\begin{align}
\label{Reg11b}
\mathcal{F}^{{\cal K},(2),\text{fin}}_{g\phi\phi} =\,&\mathcal{F}^{{\rm BPS}, (2),\text{fin}}_{f} - \zeta_3 \Big(\frac{6y(-1 + y + z)}{(1 - z)^2} \Big) 
+ 3 \zeta_2 \Big[ \frac{2y(-1 + y + z)} {(1 - z)^2} \Big( H(0;y)  
 \nonumber \\[-6mm] \nn &  
 + 4H(0;z) \Big) - \frac{H(1; y)} {z^2}(-1 + y)(-1 + y + 2z) +
     \frac{ y  H(1; z) } {z^2(1 - z)^2}  \Big(10z^2(-1 + z)  
\nonumber \\[-6mm] \nn &
+  y(-1 + 2z+ 9z^2 )\Big)
  + \frac{H(2;y)} {z^2(1 - z)^2} \Big( (-1 + z)^4 + y^2(1 - 2z + 3z^2) 
     + y(-2 
 \nonumber \\[-6mm] \nn &
 + 6z - 8z^2 + 4z^3)   \Big)  \Big]
 + \frac{3 H(0;z)}{z^2(1 - z)^2 } \Big[y H(0, 2; y)  \Big( 2 z^2 (-1 + z) + y(-1 + 2z  
 \nonumber \\[-6mm] \nn &
 + z^2)\Big)+H(1, 0; y)
  (-1 + y)(-1 + z)^2(-1 + y + 2z)  - H(2,0;y) \Big((-1 + z)^4 
\nonumber \\[-6mm] \nn &
+ y(-2+ 6z - 4z^2) - y^2(-1 + 2z + z^2)\Big)
+  H(3, 2; y) (-1 + z)^2(y^2 - z^2) \Big]
\nonumber \\[-6mm] \nn &
+ \frac{3 y H(0;y)}{z^2(1 - z)^2 }  \Big[  2z^2
(-3 + 2y + 3z)H(0, 0; z) - y(-1 + z)^2H(0, 1;z) +
 H(1, 0; z)
 \nonumber \\[-6mm] \nn &
\times \Big(2(-1 + z)z^2 + 
y(-1 + 2z + z^2) \Big)\Big] + H(1,z) \Big[ \frac{-6yH(0, 3, y)}{(1 - z)^2}(-1 + y + z) 
\nonumber \\[-6mm] \nn &
+ \frac{3H(1, 0; y)}{z^2}(-1 + y)(-1 + y + 2z) - \frac{3 H(2, 3; y)}{z^2(1 - z)^2}\Big((-1 + z)^4 + y^2(1 - 2z 
\nonumber \\[-6mm] \nn &
+ 3z^2) + y(-2 + 6z - 8z^2 + 4z^3) \Big) - \frac{3  (y^2  - z^2 )}{z^2}\Big(H(3, 0; y) - H(3, 3; y)\Big) \Big] 
 \nonumber \\[-6mm] \nn &
 - \frac{ 3 H(2;y) (-1 + y + z) }{z^2(1 - z)^2} \Big[H(0,1;z)\Big((-1 + z)^3 + y(1 - 2z + 3z^2)\Big) 
 \nonumber \\[-6mm] \nn &
 + 2 H(0, 0; z)z^2(1 + 2y - z) \Big] + \frac{3H(3,y)}{z^2} (y^2 - z^2)\Big(2H(0, 1; z) + H(1, 0; z)\Big) 
\nonumber \\[-6mm] \nn &
+ \Big(3 - \frac{3y^2}{z^2}\Big)\Big(H(3, 0, 2;y) + H(3, 2, 0; y) - H(3, 3, 2; y)\Big) -  \frac{3 H(2, 3, 2; y)}{z^2(1 - z)^2}
\nonumber \\[-6mm] \nn &
\times \Big((-1 + z)^4 + y^2(1 - 2z + 3z^2) + y(-2 + 6z - 8z^2 + 4z^3)\Big) + \frac{6yH(2, 1, 0;y)}{(1 - z)^2}
\nonumber \\[-6mm] \nn &
\times (-1 + y + z) +\frac{3(-1 + y)(-1 + y + 2z)}{z^2} \Big( H(1, 2, 0;y) + H(1,0,2; y) \Big)
\nonumber \\[-6mm] \nn &
+  \frac{6 y H(1, 1, 0; z)}{z^2(1 - z)^2} \Big( y(-1 + 2z + 3z^2)+ 4 (-1 + z)z^2\Big)- \frac{3 H(1, 1, 0; y)}{z^2}  
\nonumber \\[-6mm] \nn &
\times (-1 + y)(-1 + y + 2z)+  \frac{3 }{z^2(1 - z)^2} \Big(H(0, 1, 0; y) -    H(1, 0, 1; z)\Big) \Big(y^2(1 - 2z  
\nonumber \\[-6mm] \nn &
+ 3z^2)+ 2yz^2(-1 + z)\Big)+ \frac{6 H(0, 0, 1; z) y^2(1 - 2z)}{z^2(-1 + z)^2} 
  + \frac{3yH(0, 1, 0; z)} {z^2(-1 + z)^2}       
\nonumber \\[-6mm] \nn &
\times  \Big( 8z^2(-1 + z)+ y(1 - 2z + 7z^2)\Big)- \frac{6y(-1 + y + z)}{(1 - z)^2} H(0, 3, 2; y)
\nonumber \\[-6mm] \nn &
 + \frac{6y(-3 + 2y + 3z)}{(1 - z)^2}H(1, 0, 0; z)+ \frac{\zeta_2}{z(1 - z)^2} \Big(-10z(-1 + z)^2 + 12y^2(1 + z) 
 \nonumber \\[-6mm] \nn &
+ 3y(-3 - 4z + 7z^2)\Big)- \frac{H(0;y) H(0;z)}{z(1 - z)} \Big(-6y^2 + 14z(-1 + z) + 3y(1 + 9z) \Big) 
\nonumber \\[-6mm] \nn &
+ \frac{H(1; z) H(0; y) }{z^2} \Big(3y^2 + 3y(-1 + z) - 14z^2\Big)  - \frac{H(0; z) H(2; y)}{z(1 - z)} \Big( 3 + 6y^2 - 22z 
\nonumber \\[-6mm] \nn &
+ 19z^2 + y(-9 + 39z) \Big)- \frac{H(1; z)H(3; y)}{z^2(1 - z)} \Big( y^2(3 - 9z) + z(-3 + 8z - 5z^2)  
\nonumber \\[-6mm] \nn &
-3y(1 - 5z + 6z^2)\Big)+ \frac{6yH(0, 0; z)}{(1 - z)^2} (-1 + y + z) - \frac{H(0, 1; z)}{z^2 (1 - z)}\Big(-3y^2(-1 + z) 
\nonumber \\[-6mm] \nn &
+ 14z^2(-1 + z)+ 3y(-1 + 2z + 7z^2) \Big) + \frac{H(0, 2;y)}{z^2} \Big(3y^2 + 3y(-1 + z) 
\nonumber \\[-6mm] \nn &
- 14z^2)\Big) - \frac{H(1, 0; y)}{z^2(1 - z)} \Big( y^2(3 - 9z) + 4 z^2(-1 + z) + y(-3 + 9z)\Big) + \frac{H(1, 0; z)}{z(1 - z)^2} 
\nonumber \\[-6mm] \nn &
\times \Big(14z(-1 + z)^2 + 12y^2(1 + z) + 9y(-1 - 4z + 5z^2)\Big) + \frac{H(2, 0; y)}{z^2} \Big( 3y^2 
\nonumber \\[-6mm] \nn &
+ 3y(-1 + z) - 14z^2 \Big) - \frac{H(3, 2; y)}{z^2(1 - z)}\Big( y^2(3 - 9z) + z(-3 + 8z - 5z^2) 
\nonumber \\[-6mm] \nn &
- 3y(1 - 5z + 6z^2)\Big) + \frac{144yH(0; z)}{(1 - z)^2} (-1 + y + z)
 + \frac{12}{z(1 - z)} \Big(12y^2 
\nonumber \\[-6mm] \nn &
-12y(1 - z) +17z(1 - z) \Big) 
.
 \end{align}

The FF for the Konishi operator between $\phi\lambda\lambda$ state 
is found to be
\begin{align}
\label{Reg11c}
 \mathcal{F}^{{\cal{K}},(2),\text{fin}}_{\phi\lambda\lambda} =\,&\mathcal{F}^{{\rm BPS}, (2),\text{fin}}_{f} +\zeta_2\Big[-6H(1;y)^2 + 3H(1;z)\Big( 3H(0;y) + H(3;y)\Big)  
 \nonumber \\[-6mm] \nn &
 + \frac{9}{2}\Big(H(1;z)+H(2;y)\Big)^2- 12H(0,1;y) -9H(1,0;y) + 9H(0;y)H(2;y) 
 \nonumber \\[-6mm] \nn &
 + 3H(2,1;y)+3H(3,2;y)\Big]+3\Big(- 3H(0, 0, 0, 1; z) - 2H(0, 0, 1, 0; y) 
 \nonumber \\[-6mm] \nn &
 + 2H(0, 0, 1, 0; z) - 3H(0, 0, 3, 2; y)+ 4H(0, 1, 0, 1; y) - H(0, 1, 0, 1; z) 
 \nonumber \\[-6mm] \nn &
 - H(0, 1, 0, 2; y) + 4H(0, 1, 1, 0; y) + 3H(0, 1, 1, 0; z) - H(0, 1, 2, 0; y)  
 \nonumber \\[-6mm] \nn &
 - H(0, 3, 2, 2; y) + H(0, 3, 3, 2; y)+ H(1, 0, 0, 2; y) + 4H(1, 0, 1, 0; z) 
 \nonumber \\[-6mm] \nn &
 + H(1, 0, 2, 0; y) - 4H(1, 1, 0, 0; y) + 3H(1, 1, 0, 0; z) - 2H(1, 1, 0, 1; z) 
 \nonumber \\[-6mm] \nn &
 - 4H(1, 1, 1, 0; y) + H(1, 2, 0, 0; y) + H(2, 1, 1, 0; y) - 3H(2, 2, 1, 0; y)  
 \nonumber \\[-6mm] \nn &
 - 2H(2, 2, 3, 2; y) - H(2, 3, 0, 2; y)- H(2, 3, 2, 0; y) + 4H(2, 3, 2, 3; y) 
 \nonumber \\[-6mm] \nn &
 + 4H(2, 3, 3, 2; y) - 4H(3, 3, 2, 2; y)-4H(3, 3, 3, 2; y)\Big) -3H(0,1;z)\Big(H(2; y)^2 
 \nonumber \\[-6mm] \nn &
 +4H(2; y)H(3; y) +2H(3; y)^2\Big)+ \frac{3}{2}H(0; y)^2\Big(-\frac{3}{2}H(1; z)^2 +H(1; z)H(3; y) 
 \nonumber \\[-6mm] \nn &
 + H(0, 1; z)\Big) - \frac{9}{4}H(0; z)^2H(1; z)^2+ 3H(0, 1; z)\Big(H(0, 3; y)+2H(0;y)H(2;y)\Big) 
 \nonumber \\[-6mm] \nn &
 - 9H(0, 0; y)H(2, 2; y)+3H(1; z)^2\Big(-H(3; y)^2 + \frac{3}{2}H(0, 3; y) + 2H(3, 0; y)\Big) 
 \nonumber \\[-6mm] \nn &
 - 3H(0, 0, 1; z)\Big( 3H(2;y)+4H(3; y)\Big) - 6 H(0,1,0;y)\Big(2H(1; y) - H(2; y)\Big) 
 \nonumber \\[-6mm] \nn &
 + 3H(2; y)H(0, 1, 0; z) - 12H(3; y)H(0, 1, 1; z)+ 9H(2; y)H(1, 0, 0; y)  
 \nonumber \\[-6mm] \nn &
 - 6H(2; y)H(1, 0, 1; z)- 12H(3; y)H(1, 0, 1; z) + 6H(2; y)H(2, 1, 0; y) 
 \nonumber \\[-6mm] \nn &
 - 12H(3; y)H(2, 3, 2; y)+ H(0; z)\Big[-3H(2; y)H(1, 0; y)  
 \nonumber \\[-6mm] \nn &
+ H(1; z)\Big(3H(2; y)H(3; y) - 3H(0, 3; y)- 3H(1, 0; y) + 3H(3, 0; y) 
\nonumber \\[-6mm] \nn &
- 3H(3, 2; y)\Big) + 3H(0; y)H(3, 2; y) - 3H(0, 1, 0; y)-6H(0, 3, 2; y) 
\nonumber \\[-6mm] \nn &
+ 6H(2, 1, 0; y) + 3H(2, 3, 2; y)\Big]+H(0; y)\Big[3H(1; z)\Big(4H(3,2;y) 
\nonumber \\[-6mm] \nn &
-3H(0, 2; y)\Big)- 3H(0, 1, 0; z) + 9H(0, 1, 1; z) + 6H(0, 3, 2; y)+ 6H(1, 0, 1; z)   
\nonumber \\[-6mm] \nn &
+ 6H(2, 3, 2; y)+12H(3, 2, 2; y)\Big] + H(1; z)\Big( - 2H(3; y)^3
\nonumber \\[-6mm] \nn &
-12H(3; y)H(2, 3; y)+ 9H(0, 0, 2; y) + 3H(0, 1, 0; y) + 6H(0, 2, 3; y) 
\nonumber \\[-6mm] \nn &
+ 3H(0, 3, 0; y) - 3H(0, 3, 2; y) + 3H(0, 3, 3; y)+12H(1, 0, 0; y)   
\nonumber \\[-6mm] \nn &
- 9H(2, 0, 0; y)+ 6H(2, 0, 3; y) + 6H(2, 1, 0; y)- 6H(2, 2, 3; y)    
\nonumber \\[-6mm] \nn &
 + 3H(2, 3, 0; y) + 12H(2, 3, 3; y)- 3H(3, 0, 0; y) -12H(3, 3, 2; y)\Big)+ \frac{6z}{1-y}\zeta_3
 \nonumber \\[-6mm] \nn &
 -\frac{\zeta_2}{1-y}\Big[ 24zH(0;y) + 6z H(0,z)- \frac{3H(1;y)}{y(1-y-z)}\Big( 1-3z + 2z^2 + y^2(1+7z)  
 \nonumber \\[-6mm] \nn &
 + y(-2-4z+6z^2)  \Big)-\frac{H(1;z)}{1-y-z}\Big( 3-9z + 6z^2 + y(-3+9z)\Big)  
 \nonumber \\[-6mm] \nn &
 + \frac{H(2;y)}{y}\Big(  3-3z + y(-3+9z) \Big)\Big]+\frac{6H(0,0;y)}{1-y}\Big(  -2zH(0;z) 
 \nonumber \\[-6mm] \nn &
 + (-1+y+2z)H(1;z)  \Big) + \frac{3H(0,1;z)}{y+z}\Big[ \frac{H(0;y)}{(1-y)(1-y-z)} \Big( 2y^3 + 5y^2(-1+z) 
 \nonumber \\[-6mm] \nn &
 + z(1-3z + 2z^2)+y(3 - 6z + 5z^2)  \Big) + \frac{H(2;y)}{y(1-y)} \Big(  6y^3 + (1 - z)z + y^2(3z-7)  
 \nonumber \\[-6mm] \nn &
 + y(1 - 2z + 3z^2) \Big)-\frac{2H(3;y)}{1-y-z}\Big(  2y^2 + 5(-1 + z)z + y(-1 + 7z) \Big) \Big] 
 \nonumber \\[-6mm] \nn &
 + \frac{3H(0,2;y)}{y+z}\Big[ \frac{H(0;z)}{1-y-z}\Big( 4y^2 + 5(-1 + z)z + y(-3 + 9z)  \Big) + 2(y-z)H(1;z)  \Big]  
 \nonumber \\[-6mm] \nn &
 -\frac{3 H(1,0;y)}{y(1-y-z)}\Big[ \Big(1 + 4y^2+ 5y(-1 + z) - z\Big)\Big(H(0; z) + H(1; z)\Big) \Big]  
 \nonumber \\[-6mm] \nn &
 + H(1,0;z)\Big[ -\frac{3H(0;y)}{(1-y)(1-y-z)} \Big(  3+ 4y^2 + y(-7 + z) - z - 2z^2 \Big) 
 \nonumber \\[-6mm] \nn &
 - \frac{6z(1+z)}{(1-y)(y+z)}H(2;y)    - \frac{3H(3;y)}{y(1-y-z)} \Big(  4y^2 + 3y(-1 + z) - (-1 + z)z \Big) \Big]
 \nonumber \\[-6mm] \nn &
  + \frac{6H(1,1;z)}{y+z}\Big( (y - z)H(0, y) - (y - 3z)H(3; y) \Big)+H(2,0;y)\Big[ -\frac{3H(0;z)}{y(1-y)}
 \nonumber \\[-6mm] \nn &
\times \Big( -1 - 4y^2 + z + y(5 + z)  \Big) + \frac{6(y-z)}{y+z}H(1;z) \Big]-\frac{12z}{y+z}H(0;z)H(2,2;y)
 \nonumber \\[-6mm] \nn &
  + H(1;z)\Big[ -\frac{6(-3+3y+z)}{1-y} H(0,3;y) + \frac{3H(3,0;y)}{y(1-y-z)}
 \nonumber \\[-6mm] \nn &
\times \Big(  4y^2 + (-1 + z) (3y- z) \Big)\Big] - \frac{3H(0;z)H(3,2;y)}{y(1-y-z)}\Big( 4y^2 + (3y-z)(-1 + z) \Big) 
 \nonumber \\[-6mm] \nn &
 + \frac{3H(1;z)}{y(1-y)(y+z)}\Big[  \Big(6y^3 - (-1 + z)z - y^2(7 + z) + y(1 + 2z + 3z^2)\Big)H(2,3;y) 
 \nonumber \\[-6mm] \nn &
 + 2(-1+y)y(y-3z)H(3,2;y) \Big] - \frac{3H(1;z)H(3,3;y)}{y(y+z)(1-y-z)}\Big( (8y+z)(-1 + z)z 
 \nonumber \\[-6mm] \nn &
 + y^2(1 + 7z) \Big) - \frac{6H(0,0,1;z)}{(1-y)(y+z)(1-y-z)}\Big(y^3 -3y^2 + 2y(1 + z) + z( z^2 + z
 \nonumber \\[-6mm] \nn &
 -2)   \Big)+\frac{6H(0,0,2;y)}{1-y}(-1+y+2z) + \frac{3}{(1-y)(y+z)(1-y-z)}\Big[ \Big(  6y^3 
 \nonumber \\[-6mm] \nn &
 + y^2(-11 + 19z) + z(3 - 13z + 10z^2) + y(5 - 22z + 23z^2) \Big)H(0,1,0;y) 
 \nonumber \\[-6mm] \nn &
 + \Big(  4y^3 + 7y^2(-1 + z) + z(1 - 3z + 2z^2) + y(3 - 8z + 5z^2)  \Big) H(0,1,0;z) \Big] 
 \nonumber \\[-6mm] \nn &
 - \frac{6(y-z)}{y+z}\Big( H(0,1,1;z) - H(0,2,2;y) \Big)  +\frac{6}{1-y}\Big[ (-1+y+2z)H(0,2,0;y) 
 \nonumber \\[-6mm] \nn &
 - (-3+3y+z)H(0,3,2;y) - (-1+y+4z)H(1,0,0;y) \Big] 
 \nonumber \\[-6mm] \nn &
 - \frac{3H(1,0,1;z)}{(1-y)(y+z)(1-y-z)}\Big(  6y^3 + y^2(-11 + 7z) + z(-1 - z + 2z^2) + y(5 
 \nonumber \\[-6mm] \nn &
 - 6z+ 3z^2) \Big) - \frac{3H(1,0,2;y)}{y(1-y-z)}\Big( 1 + 4y^2 - 5y(1 - z) - z \Big) + \frac{3H(1,1,0;y)}{y(1-y)(1-y-z)}
 \nonumber \\[-6mm] \nn &
\times \Big( 1 - 3z + 2z^2 + y^2(1 + 7z) + y(-2 - 4z + 6z^2)  \Big)+ \frac{6H(1,1,0;z)}{(1-y)(y+z)(1-y-z)}
 \nonumber \\[-6mm] \nn &
\times \Big(  (y^2+z^2)(-1 + z) + y(1 - z + 2z^2) \Big)-\frac{3H(1,2,0;y)}{y(1-y-z)}\Big( 1 + 4y^2 + 5y(-1 + z) 
 \nonumber \\[-6mm] \nn &
 - z \Big) + \frac{6(-1+y+2z)}{1-y}H(2,0,0;y)+\frac{6(y-z)}{y+z}\Big( H(2,0,2;y)+ H(2,2,0;y) \Big) 
 \nonumber \\[-6mm] \nn &
 - \frac{6(3y+2z)}{y+z}H(2,1,0;y)+ \frac{3H(2,3,2;y)}{y(1-y)(y+z)}\Big(  6y^3 - (-1 + z)z - y^2(7 + z) 
 \nonumber \\[-6mm] \nn &
 + y(1 + 2z + 3z^2) \Big)+ \frac{3}{y(1-y-z)}\Big(    4y^2 + (3y-z)(-1 + z)\Big) \Big(H(3,0,2;y) 
 \nonumber \\[-6mm] \nn &
 + H(3,2,0;y) \Big)-\frac{6(y-3z)}{y+z}H(3,2,2;y) - \frac{3H(3,3,2;y)}{y(y+z)(1-y-z)}\Big(  (-1+z)(8 yz
 \nonumber \\[-6mm] \nn &
 +z^2) + y^2(1+7z) \Big) - \frac{1}{1-y} \Big(2\zeta_2(5 - 5y + 6z)+(8 - 8y+ 12z)H(1,0;y)\Big)  
\nonumber \\[-6mm] \nn &
+ 14H(0; y)H(0; z)+22H(0; y)H(1; z) - 14H(0; z)H(2; y) 
\nonumber \\[-6mm] \nn &
-\frac{1}{y+z}\Big[4(2y+5z)\Big(H(1; z)H(3; y)+H(3,2;y)\Big) + (22y + 34z)H(0, 1; z) \Big] 
\nonumber \\[-6mm] \nn &
+ 22H(0; y)H(2;y)  - 10H(1, 0; z) -\frac{108z}{1-y} H(0;y) + \frac{108z}{y+z}\Big(H(1;z)+H(2;y)\Big)
\nonumber \\[-6mm] \nn &
+204 
.
 \end{align}

\bibliography{master} \bibliographystyle{utphysM}
  
\end{document}